\documentclass[aip,pop,reprint]{revtex4-1}

\usepackage{amsmath}
\usepackage{amssymb}
\usepackage{graphicx}
\usepackage{color}
\usepackage{bm}
\usepackage{dcolumn}

\begin{document}

\title{New radiographic image processing tested on the simple and double-flux platform at OMEGA}
\author{\surname{Olivier} Poujade}
\email[]{olivier.poujade@cea.fr}
\affiliation{CEA, DAM, DIF, F-91297 Arpajon, France}
\author{\surname{Michel} Ferri}
\affiliation{CEA, DAM, CESTA, 15 Avenue des Sablieres, F-33114 Le Barp, France}
\author{\surname{Isabelle} Geoffray}
\affiliation{CEA, DAM, Valduc, F-21110 Is-sur-Tille, France}

\date{\today}

\begin{abstract}

Ablation fronts and shocks are two radiative/hydrodynamic features ubiquitous in inertial confinement fusion (ICF). A specially designed shock-tube experiment was tested on the OMEGA laser facility to observe these two features evolve at once and to assess thermodynamical and radiative properties. It is a basic science experiment aimed at improving our understanding of shocked and ablated matter which is critical to ICF design. At all time, these two moving ``interfaces'' separate the tube into three distinct zones where matter is either ablated, shocked or unshocked. The {\it simple-flux} or {\it double-flux} experiments, respectively one or two halfraum-plus-tube, have been thought up to observe and image these zones using x-ray and visible image diagnostic. The possibility of observing all three regions at once was instrumental in our new radiographic image processing used to remove the backlighter background otherwise detrimental to quantitative measurement. By so doing, after processing the radiographic images of the 15 shots accumulated during the 2013 and 2015 campaigns, a quantitative comparison between experiments and our radiative hydrocode simulations was made possible. Several points of the principal Hugoniot of the aerogel used as a light material in the shock-tube were inferred from that comparison. Most surprisingly, rapid variations of relative-transmission in the ablated region were observed during radiographic irradiations while it remained constant in the shocked region. This effect might be attributed to the spectral distribution variability of the backlighter during the radiographic pulse. Numerically, that distribution is strongly dependent upon NLTE models and it could potentially be used as a mean to discriminate among them.

\end{abstract}

\maketitle

\section{Introduction}

Hydrodynamic shock-tubes (HST) have been used for more than 100 years \cite{Fomin2010} and are still in use \cite{Biamino2015, Zhigang} to assess the physics of shock-waves on matter. However, the densities, pressures and temperatures accessible to these experiments are limited. In supersonic flows encountered in astrophysics or in fusion experiments, thermodynamics conditions can be more drastic. In order to meet these requirements, laser shock-tube (LST) experiments have been developed over the last thirty years on many high power laser facilities such as NOVA \cite{lstnova}, OMEGA \cite{lstomega} and NIF \cite{lstnif}.

\par

On one hand, measurements on HST can be carried out with precise diagnostics, resolved in space and time, able to measure flow velocities everywhere in the field of view (with particle image velocimetry \cite{piv}) which is a precious information when it comes to inferring turbulent quantities in order to adjust turbulent mixing models parameters for instance. This is made possible by the fact that characteristic size of these tubes is of order of a few meters, sections are of the order of tens of centimeters and the hydrodynamics evolved on time scales of a few milliseconds. On the other hand, HSTs suffer from the limited range of shock intensities. Apart from a small number of experiments using explosives, most HSTs rely on a high pressure gas chamber \cite{Fomin2010} to generate a shock. They are designed in such a way that only one shock is triggered for each shot. A reshock is produced by the rebound of the initial shock on a surface at the other end of the shock-tube \cite{}. It is often used to study the interaction of a shock wave with a developed mixing zone but the intensity of this return-shock cannot be tuned. 

\par 

In LST experiments, on the other hand, the high pressure compartment, or high power explosive, is replaced by high power laser beams hitting a piston (directly or indirectly) producing a sufficiently high ablative pressure to launch a shock propagating through the material enclosed in the tube. If a similar laser drive is used at the other end of that tube, it is possible to adjust the timing and intensity of a second counter-propagating shock at any level. The pulse shaping is a clear advantage of LSTs over HSTs. In addition to that, since the element inside the tube are solid, it is technically possible to give the interface any shape we think fit without using a membrane or any other artifact that could affect the supersonic flow to be observed (with splinters). This kind of platform can then be used to study Richtmyer-Meshkov \cite{lstRM}, Rayleigh-Taylor \cite{lstRT}, Kelvin-Helmholtz \cite{lstKH} instabilities and many others such as the interaction of a shock with a turbulent mixing zone. However, the millimeter-sized platform along with the nanosecond time scale hydrodynamics make a joint space and time resolved observation difficult. 

\par 

In this article, one presents the design of a LST experiment shot on the OMEGA laser facility. The purpose was to test a concept (material, geometry, etc) that could be used in the future as a starting point for the design of a similar experiment on LMJ. The platform consists of one (or two) halfraum(s) located at one (or each) end(s) of a tube (in section \ref{sec:exp}). Two campaigns have been carried out on the OMEGA laser facility at Rochester (USA) in 2013 and 2015. The first campaign in 2013 was devoted to the one-halfraum-plus-tube configuration called {\it simple-flux} throughout this article. The last campaign, in 2015, was focused on improving the {\it simple-flux} configuration (choice of materials, geometry) and testing the two-halfraum-plus-tube design, referred to as {\it double-flux} in this article. These platforms will be used to benchmark radiative hydrocodes in the shocked regime but also in the underdense regime which is not easy to characterized experimentally, while it may have important effects on the hydrodynamics \cite{poujbonvdb}.

\par

Materials in the tube were imaged with an x-ray framing camera (XRFC) and shocks were tracked using a streaked optical pyrometer (SOP) in an unconventional manner (side-on). In both cases, the experimental results were compared with simulations (described in section \ref{sec:sim}). Many efforts have been made on data processing of high energy density experiments \cite{drake, abeltransf} in the literature to get the most quantitative measurement out of noisy images. Here, the radiographic images (from XRFC) provided us with a clear view on shocks and ablation fronts with their positions and shapes. A quantitative study of intensity profiles of these radiographic image was made possible by a new analysis process that will be thoroughly described in section \ref{sec:remov}. A consequence of that process is that simulations can readily be compared to experimental intensity profiles. As a result, it provided us with the possibility of making precise measurements of density profiles (through Abel transformations) in the tube and with the possibility of measuring pressures along the principal Hugoniot\footnote{Locus of achievable shocked state from a given initial unshocked state} of the material crossed by the shock. in order to make necessary corrections on our equations of state for the shock dynamic to agree with experimental data collected from the SOP. Unexpectedly, it also provided us with information on the importance and quick variation of the low energy \mbox{($<$ 1 keV)} part of the backlighter intensity spectrum during the course of its pulse compared to the $K_\alpha$ main line emission (usually $>$ 4 keV). Numerically, this effect is sensitive to NLTE models and the experimental results could be used to discriminate these models.

\begin{figure}[h] 
\includegraphics[width=8cm]{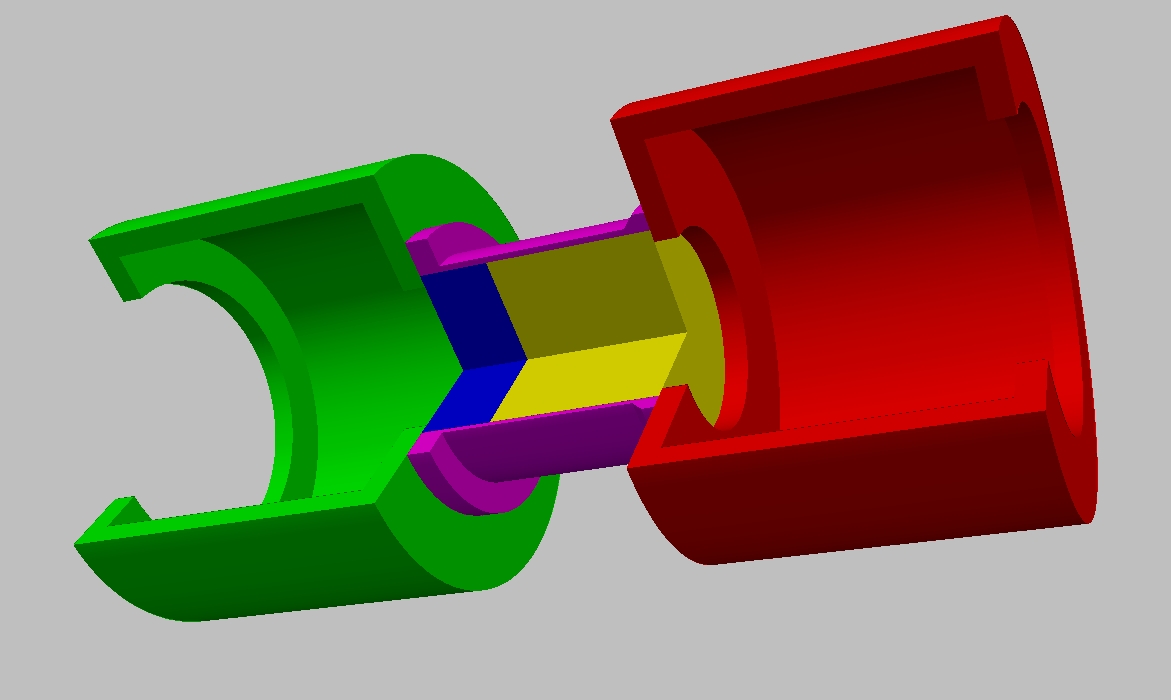}
\caption{Geometry of the full {\it double-flux} target. In green and red are the two halfraums used for x-rays conversion. The wall of the shock-tube is in purple. The piston (in blue) can be 0, 150 or 300 $\mu$m thick and the aerogel, in yellow, fills up the rest of the tube. The {\it simple-flux} experiment has the same geometry without the red halfraum.}\label{fig:geo}
\end{figure}

\section{\label{sec:exp} Experimental setup}

\begin{figure}[h] 
\includegraphics[width=8cm]{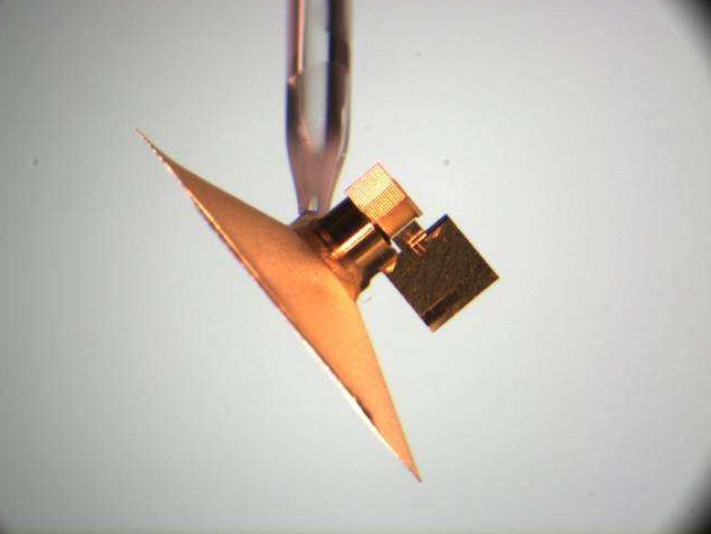}
\includegraphics[width=8cm]{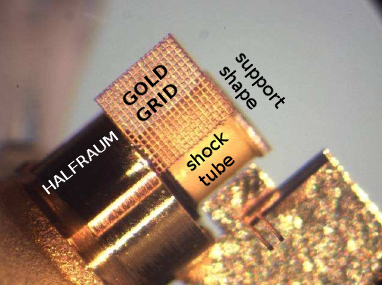}
\caption{Actual pictures of the simple shock platform as shot in 2013 and 2015 (fiducial grid is smaller in 2015, 700 $\mu$m instead of 1000 $\mu$m). (Top) general view of the platform. The gold halfraum cylinder, converting the laser irradiation into x-rays, is just above the cone shaped gold washer. The smaller cylinder is the shock-tube on top of the halfraum. (Bottom) close-up view of the halfraum and the shock-tube as seen from the radiography line of sight with the gold grid.}\label{fig:tr1}
\end{figure}

\begin{figure}[h] 
\includegraphics[width=7cm]{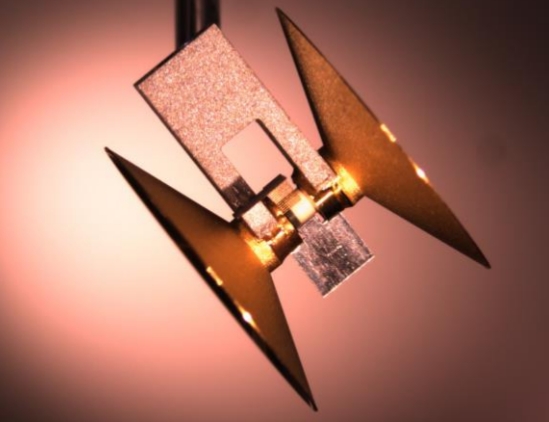}
\includegraphics[width=7cm]{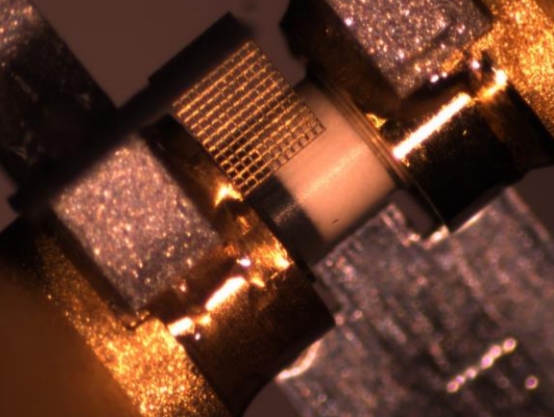}
\caption{(Color online) Actual pictures of the double-shock platform as shot in 2015. (Top) general view of the platform with two halfraums at both ends of a shock-tube. (Bottom) close-up view of the shock-tube with the ``heavy'' piston in black polyimide, on the left hand side of the tube, and the ``light'' white DVB aerogel, in the remaining part of the tube.}\label{fig:tr2}
\end{figure}

The design of a laser shock-tube experiment is presented in this section. Two distinct platforms were experimented on the OMEGA high energy laser facility. They are called {\it simple-flux} and {\it double-flux} depending upon the number of x-rays conversion halfraums involved (see Fig.\ref{fig:geo}). The {\it simple-flux} platform (see Fig.\ref{fig:tr1} after fabrication) is dedicated to single shock generation. The {\it double-flux} platform, on the other hand, (see Fig.\ref{fig:tr2} after fabrication) is devoted to shock and reshock experiments.

\par 

The first campaign was shot in september 2013 on OMEGA in {\it simple-flux} configuration. It was dedicated to testing materials and diagnostics for the {\it double-flux} campaign of 2015: a streaked optical pyrometer (SOP) and an X-ray framing camera (XRFC). 

\par

The second campaign was shot in september 2015 on OMEGA: in {\it simple-flux} configuration for shots 78804 to 78807 and in {\it double-flux} configuration for shots 78808 to 78811. 

\par 

Overall, 15 shots were carried out in 2013 and 2015. Differences and results are summarized in table \ref{tab:table1}. There were several differences: some related to the geometry ($d_\mathrm{ab}$: the piston thickness) and type ($\mathrm{sf}$: {\it simple-flux} and $\mathrm{df}$: {\it double-flux}), some related to the materials themselves (piston made out of polystyren: PS, or polyimid: PI), some related to diagnostic settings (BL: backlighter material, ND: neutral density used on the SOP diagnostic and $t_\mathrm{radio}$: the time of the radiographic image).

\begin{table}[h]
\caption{\label{tab:table1}%
List of shots considered in this article with their essential characteristics : shot identification number (ID), thickness of the ablator ($d_\mathrm{ab}$), type of the experiment -- {\it simple-flux} or {\it double-flux} -- (type), material of the backlighter (BL), number of images selected among the 16 images of the plate ($N_\mathrm{img}$) and the time of radiography ($t_\mathrm{radio}$ in ns) which is given at $\pm$ 0.2 ns}
\vspace{0.3cm}
\begin{ruledtabular}
\begin{tabular}{lccccccccc}
ID&$d_\mathrm{ab}$&type&BL& $N_\mathrm{img}$ & $t_\mathrm{radio}$ (ns)& $z_s$ & SOP& piston& ND\\
\hline
 & & & & & &\\
70989\footnote{campaign 2013}&300&$\mathrm{sf}$&V&5&11& 550 & $\mathrm{no}$&PS&-\\
70990&300&$\mathrm{sf}$&V& 4&14& 660 &$\mathrm{yes}$&PS&1.5\\
70991&150&$\mathrm{sf}$&Fe& -&11& $-$ &$\mathrm{no}$&PS&-\\
70992&150&$\mathrm{sf}$&Fe& -&8& 460 &$\mathrm{yes}$&PS&0.8\\
70993&150&$\mathrm{sf}$&V& 3&8& 480 &$\mathrm{yes}$&PS&0.8\\
70994&150&$\mathrm{sf}$&Fe& -&7& 470 &$\mathrm{no}$&PS&-\\
70995&150&$\mathrm{sf}$&Fe& -&6& 430 & $\mathrm{no}$&PS&-\\
78804\footnote{campaign 2015}&300&$\mathrm{sf}$&Ti&5&14 & 610 &$\mathrm{yes}$&PI&0.0\\
78805&0&$\mathrm{sf}$&Ti&5&5 & 470 &$\mathrm{yes}$& -&0.0\\
78806&300&$\mathrm{sf}$&Sc&5&14 & 660 & $\mathrm{yes}$&PI&0.0\\
78807&0&$\mathrm{sf}$&Sc&6&7 & 680 &$\mathrm{yes}$&-&0.0\\
78808&300&$\mathrm{df}$&Ti& 4&14& 580 \footnote{colliding shocks}&$\mathrm{yes}$&PI&0.0 \\
78809&300&$\mathrm{df}$&Ti& -&19& $-$& $\mathrm{yes}$&PI&0.0 \\
78810&300&$\mathrm{df}$&Ti& -&24&  $-$ &$\mathrm{yes}$&PI&0.0 \\
78811&300&$\mathrm{df}$&Ti& -&17& $-$ &$\mathrm{yes}$&PI&0.0 \\
\end{tabular}
\end{ruledtabular}
\end{table}

\subsection{\label{sec:st} The shock tube}

The shock tube is 1 millimeter long with an inner diameter of 900 microns and a 25 microns thick wall made up of translucent epoxy resin with composition $C_{41.5}\,H_{48.9}\,O_{8.3}\,N_{1.3}$ (in $\%$at). The inner part is composed of two sections. The heavy piston is made up of polystyrene (PS, with $\rho=$1.046 g/cm$^3$), in 2013, or polyimide (PI, with $\rho=$1.414 g/cm$^3$), in 2015, with thickness $d_\mathrm{ab}$=0, 150 or 300 microns (depending on shot number). The rest of the 1000 micron shock-tube is filled with an organic aerogel (DVB, with $\rho=$0.3 g/cm$^3$) whose composition is $C_{48.6}\,H_{51.12}\,O_{0.22}\,N_{0.05}\,S_{0.01}$ (in $\%$at). 

\par 

In both campaigns, the piston was undoped. It makes the transition between ablated and non-ablated material visible on the radiographic images. This proved to be very useful for the quantitative analysis of these images but an extensive discussion of this point is deferred to section \ref{sec:remov}. The drawback is that the contrast at the interface between the shocked piston and the shocked aerogel is small, making that interface between heavy and light material unreliably difficult to observe (in order to examine an instability development for instance). If the important feature to watch laid at the interface, the solution would consist in doping the whole piston or a tracer strip \cite{drakestrip} to make the evolution of its shapes visible. This would certainly come at the expense of the ablation front that would not show up in this condition.

\subsection{Indirect drive and laser pulse}

The same empty gold-halfraum is used on both {\it simple-flux} and {\it double-flux} platforms. The walls are 100 $\mu$m thick. The laser entrance hole (LEH) is 1200 $\mu$m in diameter whereas the shock-tube entry hole is 800 $\mu$m in diameter. The geometry of the halfraum was designed in such a way that a peak radiative temperature of about 200 eV could be reached with a nanosecond laser pulse of 7 kJ (with 14 beams of 500J each of the OMEGA laser). The first halfraum to be irradiated, to create the first shock on the piston side of the tube, will be called X1. The other halfraum, irradiated with a delay to create the reshock, will be called X2 throughout this article. In the case of a {\it simple-flux} experiment only X1 is attached to the shock-tube.

\par

Beam configuration was not a straightforward issue on OMEGA for the {\it double-flux} platform. Three sets of beams are needed : one for each halfraum and one for the backlighter. In the best case scenario, these three sets of beams should  be controlled in time independently of each other. Luckily, there are three pilots on OMEGA, each commanding one leg (a cluster of 20 beams). Unfortunately, all of the 20 beams of any set cannot be made to point at the same location (interesting locations being the two LEHs or the backlighter). Pointing directions of the beams associated to different SSD pilots are entangled over the 4$\pi$ steradians of the spherical target chamber. A trade-off between the power that gets to these locations and the ability to separately control the timing of the beams needs to be made because a delay between the laser pulses of the two halfraums needs to be set (for shock and reshock to converge at a compulsory location and time) just as between the halfraums and the backlighter (to set the time of the radiography).  

\par

In the end, the beams were distributed according to the following list. On leg1, 6 beams of 450 J each are used for the backlighter (a total amount of 2.7 kJ). Legs 2 and 3 remain to irradiate halfraums X1 (and X2 in the case of a {\it double-flux} experiment). The shock-tube axis is positioned along the P6-P7 axis of the OMEGA target chamber. Exactly 6 beams from leg2 point toward X2 (P6 side) and 8 beams point toward X1 (P7 side). An additional 8 beams from leg3 point toward X2 and 7 beams point toward X1. An overall 14 beams of 500 J point toward X2 (a total of 7 kJ) and 15 beams of 500 J point toward X1 (a total of 7.5 kJ). Beams devoted to the backlighter are on the same leg and can be controlled independently. Therefore, the time of the radiography can be set without any constraint. On the contrary, beams to X1 and X2 cannot be controlled independently. Both pilots (for leg 2 and 3) have to be triggered at once but the {\it double-flux} experiment requires that the beams to X2 should be delayed with respect to those to X1 (in order for the shock and reshock to be delayed accordingly). That chain of events is made possible by increasing the optical path of beams attributed to X2. It can be done by moving mirrors in the corresponding laser chain allowing a maximum delay of 10 ns. The actual {\it double-flux} shots were carried out using a 9.5 ns delay.

\par

\begin{figure}[h] 
\includegraphics[width=8cm]{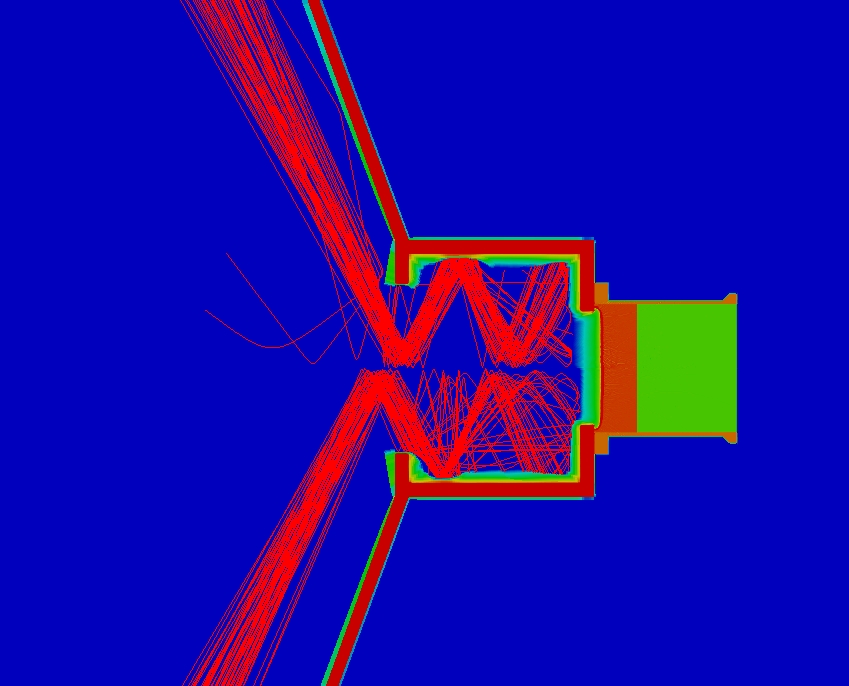}
\caption{Laser beam, halfraum, cone washer and shock tube at t=0.9 ns, during the nanosecond pulse. (Top) pointing shifted 100 $\mu$m closer to the halfraum than the nominal pointing, (bottom) pointing shifted 100 $\mu$m farther from the halfraum than the nominal pointing. The hydrodynamic motions in the tube have been proven to be unaffected by these pointing changes as long as the drive remains indirect, that is to say, as long as the laser ends its path on the gold wall around the tube entry hole.}\label{fig:point}
\end{figure}

Beams selected on leg 2 and 3 get in the halfraums with an acute angle of 63.5 degrees. The actual pointing of the beams on the axis of the experiment was 100 $\mu$m outside the LEH. One does not have free hands to choose the pointing location on the axis. The choice made provides two reflexion in the cavity and a small fraction of direct laser interaction with the piston. The size of the halfraum is constrained by the angle of irradiance of the laser beams. Since the drive must remain indirect, the design must be done in such a way that, after reflexions on the wall of the halfraum, the laser should not end its path on the piston. It should be reflected back by the vertical wall delimiting the tube entry hole. We have made a simulation of what would happen if the pointing location was slightly altered. Simulations with $\pm$100 $\mu$m pointing-shift with respect to nominal pointing, proved the indirect drive was quite robust to pointing uncertainties (see Fig.\ref{fig:point}).

\par

As long as the pointing is such that the beams do not hit the piston directly, the chronometry of shocks and of ablation fronts does not depend on the pointing. It mostly depends, for a nanosecond pulse, upon the quantity of energy that gets in the halfraum (7 kJ total in our case). On the contrary, if the pointing is such that a part of this energy hits the piston directly, the chronometry can be drastically different depending essentially on the balance between direct and indirect drive in the irradiation of the piston. The agreement between the simulated chronometry of the shock and of the ablation front with its experimental counterpart (from XRFC images and SOP as we will show later on in this article) gives us good confidence that the balance direct/indirect was close to zero.

\par

Finally, the amount of backscattered light from the halfraum was monitored during the pulse. As expected, since these halfraums are empty, the stimulated Brillouin scattering (SBS) and stimulated Raman scattering (SRS) signals were negligible. Therefore, it is safe to conclude that the coupling between laser beams and halfraums is close to perfect.

\subsection{Diagnostic}

For each shot, the shock-tube is imaged with a side-on radiography using the x-ray framing camera (XRFC). The x-ray source is a foil (backlighter), 3000 $\mu$m in diameter and 10 $\mu$m thick, 4000 $\mu$m away from the tube, made up of a choice of iron (with a $K_\alpha$ emission line at 6.0 keV), vanadium ($K_\alpha$ emission at 5.2 keV),  titanium ($K_\alpha$ emission at 4.7 keV) or scandium ($K_\alpha$ emission at 4.3 keV). The backlighter is irradiated with a 2.7 kJ nanosecond-pulse made of 6 beams of 450 J at 3$\omega$ (350 nm). For each shot, the experimental results of the XRFC is a set of 16 radiographic images taken within 50 ps of one another. More detail on XRFC can be found in section \ref{sec:xrfc}.

\par

\begin{figure}[h] 
\includegraphics[width=6cm]{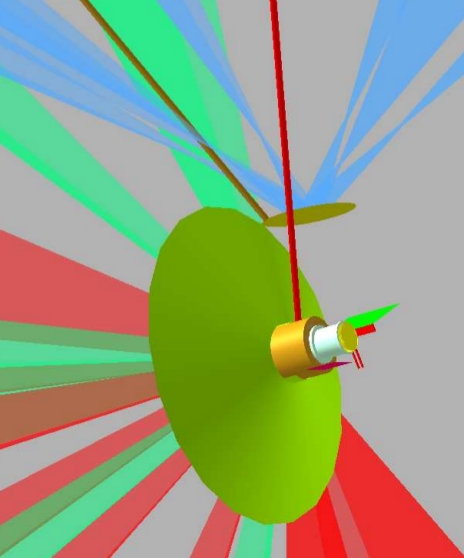}
\caption{Beam configuration from the three different pilots (in red and green, leg 2 and 3, to the halfraum) and in blue, leg 1, to the backlighter.}\label{fig:bea}
\end{figure}

Iron and vanadium  backlighters where used in 2013, for shots labelled from 70989 to 70995. The iron, with a harder spectrum than vanadium (6.0 keV instead of 5.3 keV), was expected to let more radiation go through the tube. With the uncommon shallow drive on the backlighter ($\approx 60$ degrees with respect to the perpendicular to the BL due to beam restrictions, see Fig.\ref{fig:bea}) it was not sure what absolute intensity to expect on the radiographic plate and iron was, among all the backlighters, the safest in terms of radiation transmission. It turned out that, even in this particular configuration, the intensity was not an issue and contrast was, of course, better with a softer spectrum such as vanadium. 

\par 

During the 2015 campaign, titanium and scandium backlighters, with even softer x-ray spectrum emission, where used to increase contrast between shocked and unshocked region. The material of the piston was also changed in the 2015 campaign from polystyren (PS with $\rho$=1 g/cm$^3$) to polyimid (PI with $\rho$=1.4 g/cm$^3$) for that reason. The increase in density goes in the sense of a contrast increase but it also modifies the dynamics of the piston with more inertia.

\par 

In addition to side-on radiographic image, the streaked optical pyrometer (SOP) was also turned on during the experiments. It was also pointing side way in order to collect the visible (and near infrared) light emitted by the shock-tube wall itself or by the shocked plastic inside the tube, allowing to resolve the position of the shock, and any emitting part, in time. Its sensitivity to visible light requires shielding from any source of luminosity in this particular spectral domain which includes laser beams themselves (wavelength of 351 $\mu$m) and any plasma flowing out of the halfraum. This is the reason why a gold cone shaped washer is displayed around the laser entrance hole (LEH), see Fig.\ref{fig:tr1} and \ref{fig:tr2}, in order to prevent any dispersed laser light from reaching the shock tube directly and being collected by the telescope bringing the visible light outside the target chamber to be analyzed by the SOP. Indeed, the laser beams converging to the center of the target chamber are made mainly of 3$\omega$ laser light but some unfiltered 2$\omega$ laser light can go through. A 2$\omega$ and a 3$\omega$ beam in the target chamber look alike but the focal point of the 2$\omega$ is shifted by approximately 1 cm ahead of the 3$\omega$ focal point. Another way to put it is that on a fixed plane of observation (perpendicular to the beams propagation), the 2$\omega$ beam spread on a bigger surface than does the 3$\omega$ beam. As a consequence, it is not impossible to get some 2$\omega$ reflecting off the surface of the shock-tube and/or preheating the surface of the shock-tube. Either way, it is possible that 2$\omega$ can pollute the SOP images during the 1 ns laser pulse. More detail on the SOP will be found in section \ref{sec:sop}.

\section{\label{sec:sim} Simulations}

\subsection{Setup}

Full 2D axisymmetric integrated simulations of both platforms ({\it simple-flux} and {\it double-flux}) have been carried out with the radiation transfer code FCI2 \cite{fci2}. SESAME equations of state \cite{SESAME} and SCO opacity \cite{SCO} tables are used. The important part of the actual geometry of both platforms is axisymmetrical apart from few features that do not affect the functioning of the platform such as the fiducial grid or the shield for the SOP. 

\par 

However, the laser irradiation is 3D (cf. Fig. \ref{fig:bea}) in the actual experiment. In our simulations, the irradiation is axisymmetrized with the accurate cone angle of 63 degrees and laser rays with wavelength $\lambda=350$ $\mu$m. The pointing is on axis, $100$ $\mu$m off the LEH as in the experimental configuration.

\par 

\begin{figure*}[t] 
\includegraphics[width=4cm]{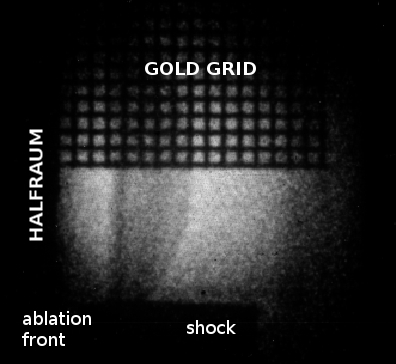}\hspace{.2 cm}\includegraphics[width=4cm]{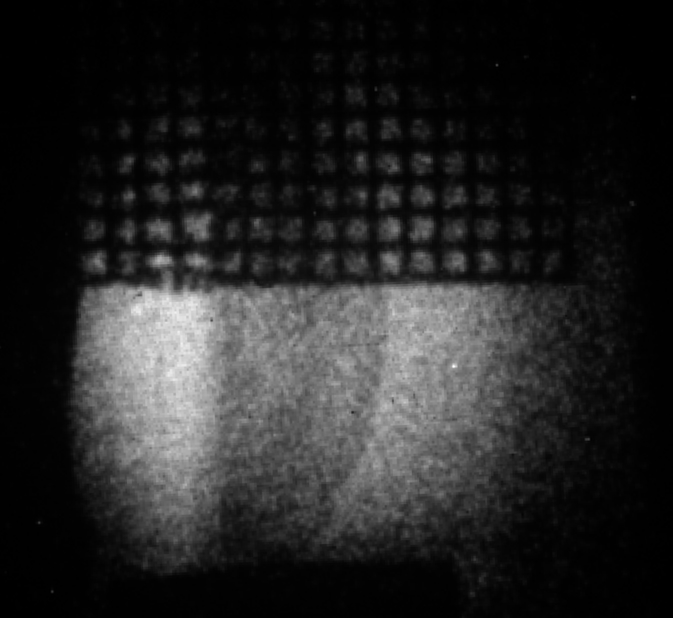}\hspace{.2 cm}\includegraphics[width=4cm]{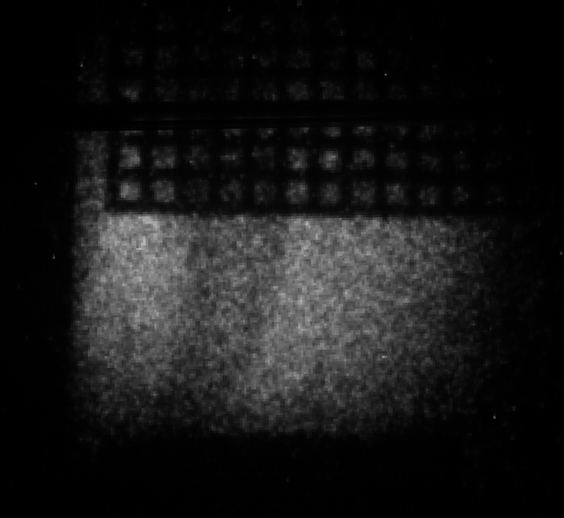}\hspace{.2 cm}\includegraphics[width=4cm]{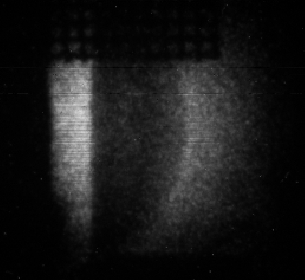}\\
\vspace{.1cm}
\includegraphics[width=4cm]{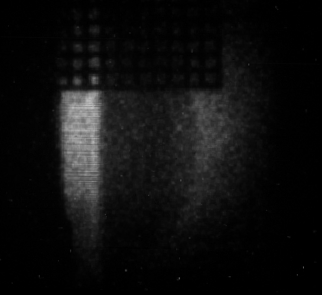}\hspace{.2 cm}\includegraphics[width=4cm]{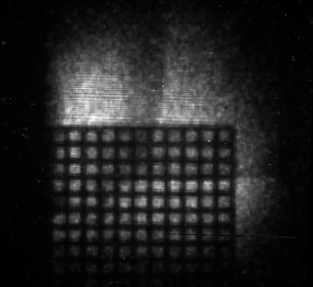}\hspace{.2 cm}\includegraphics[width=4cm]{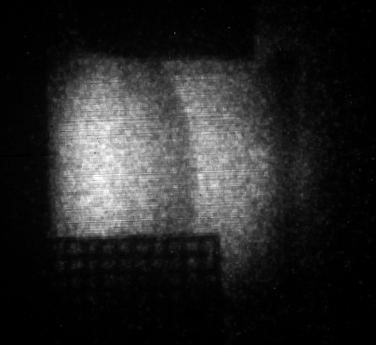}\hspace{.2 cm}\includegraphics[width=4cm]{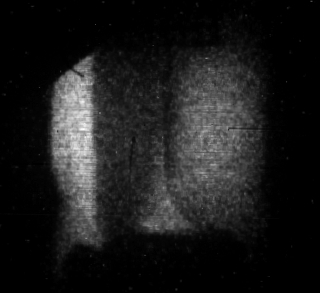}
\caption{Side-on radiographic images with XRFC on OMEGA. (Top, from left to right) shot 70989 (simple-flux, with piston 300 $\mu$m in PS), shot 70990, shot 70993 (simple-flux, with piston 150 $\mu$m in PS) and shot 78804 (simple-flux, with piston 300 $\mu$m in PI). (Bottom, from left to right) shot 78806 (simple-flux, with piston 300 $\mu$m in PI), shot 78805 and 78807 (simple-flux, without piston) and shot 78808 (double-flux).}\label{fig:rad}
\end{figure*}

A pointing offset has an effect on the distribution of the ion and electron temperature within the halfraum but the radiative temperature balances rapidly in such a small cavity. Its distribution is almost uniform at any given instant in the halfraum, away from the LEH. This is the reason why 3D effects were assumed to be small when it comes to the hydrodynamic evolution observed. This is corroborated by the observation of the ablation front in Fig.\ref{fig:rad} which is a straight line perpendicular to the axis of the tube within the precision of the radiographic images. If significant non-uniformities in the flux were actively involved, the ablation front would be distorted (imprint). Therefore, as far as the hydrodynamics of the shock-tube is concerned, it is considered as good as axisymmetric. 

\par

Many technical features, such as the reinforcement of the tube at both end or the shielding cone(s) have been taken into account because, even if they do not affect much the hydrodynamics in the tube, they can affect post-processing simulation of the radiographic images. 

\subsection{\label{sec:hydro} Hydrodynamic evolution in the shock-tube}

\begin{figure}[h] 
\includegraphics[width=8cm]{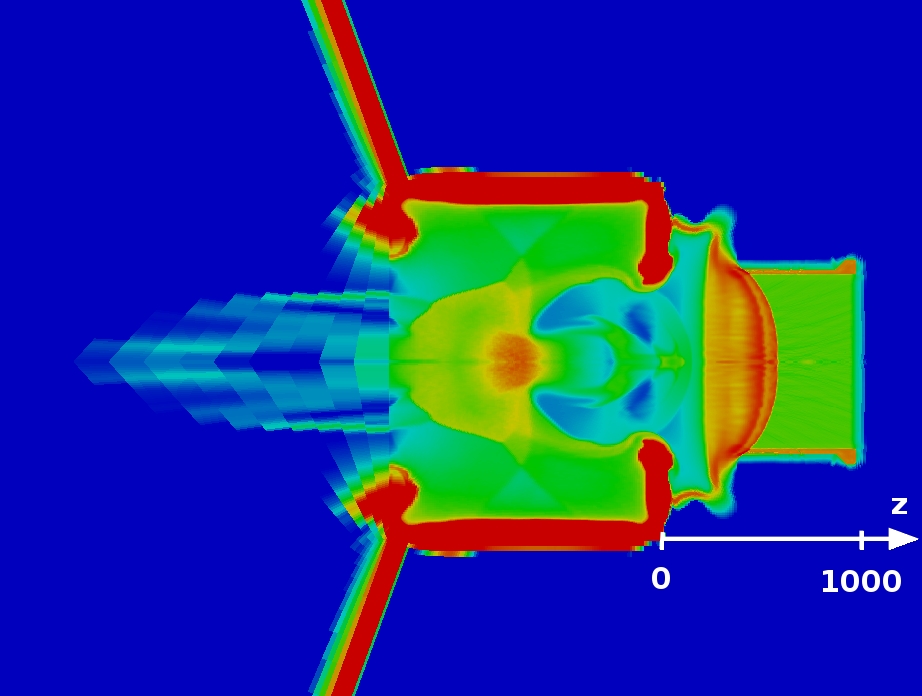}
\caption{(color online) Color plot of the density in a simple-flux platform at t=14 ns. The shock is curved in the tube and the ablation front is straight and perpendicular to the axis. The position $z$=0 $\mu$m correspond to the end of the tube in contact with the halfraum and $z$=1000 $\mu$m corresponds to the other end of the tube.}\label{fig:dens}
\end{figure}

At $t=$0 ns, laser beams irradiate the halfraum walls transforming laser light into a bath of x-rays that fills up the cavity. The spectrum of these x-rays is made up of a blackbody distribution (a large part of the radiation thermalized in the cavity) along with a significant M-band from gold. In turn, these x-rays ablate the outer surface of the piston where the ablation pressure increases drastically as $\approx T_r^{3.5}$. It triggers a curved shock (see Fig.\ref{fig:dens}) that travels across the piston and across the aerogel with Mach numbers ranging from 10 to 20. The curvature of such shock has also been observed before and authors\cite{lstRT} made a list of possible causes concluding that it was due to the ``imprint of the experimental condition''. Our analysis concurs with this conclusion : the curvature of the shock mainly depends upon the radius of the tube-entry hole. The smaller the radius, the more curved the shock. These conclusions are inferred from a series of simulations where the radius of the tube entry hole was varied.

\par

\begin{figure}[h] 
\includegraphics[width=8cm]{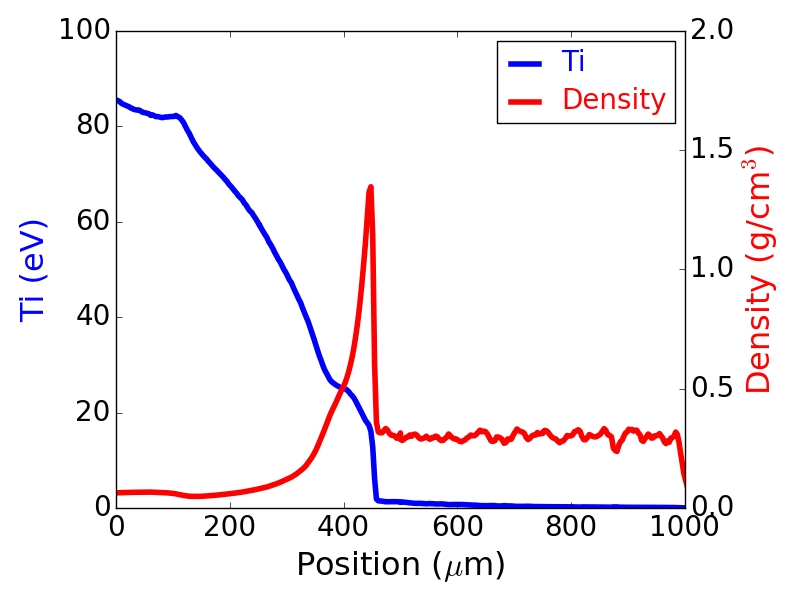}\\
\includegraphics[width=8cm]{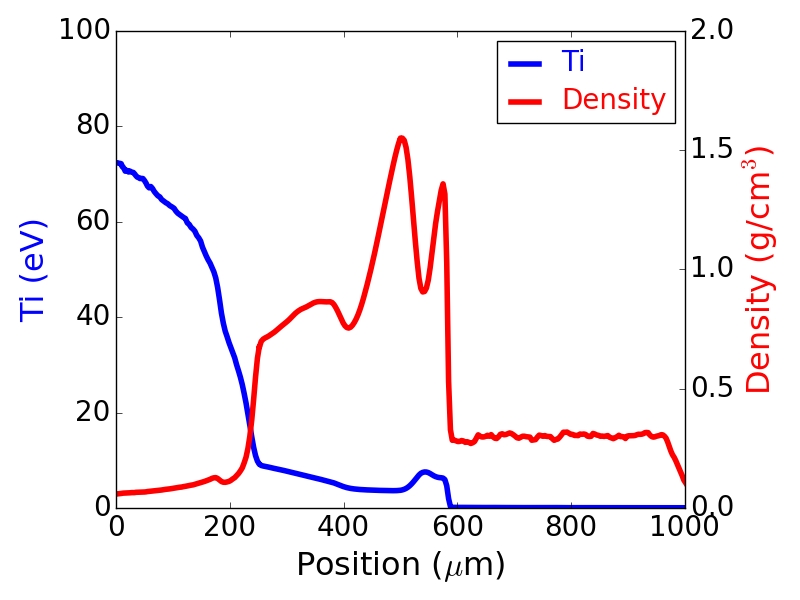}
\caption{(Color online) Ion temperature and density profiles for shot 78805 (top) and 78804 (bottom).}\label{fig:TIRO}
\end{figure}

\begin{figure}[h] 
\includegraphics[width=8cm]{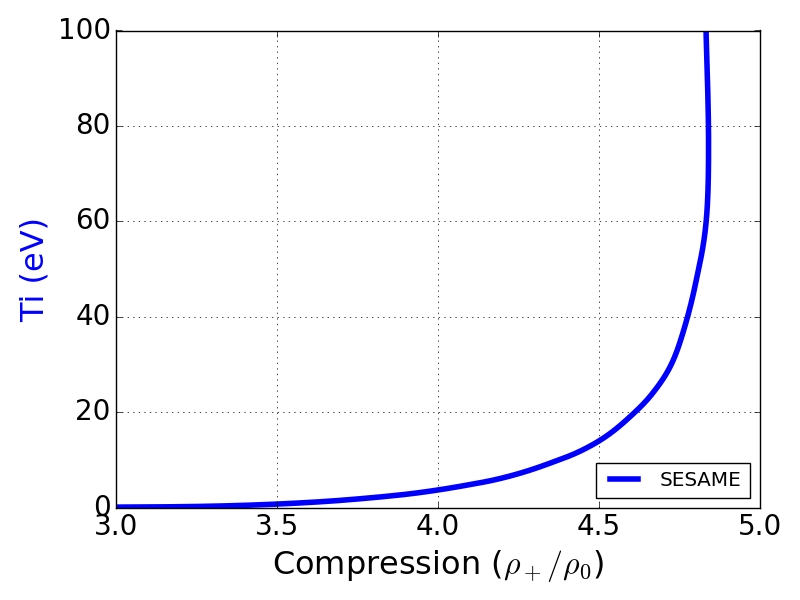}
\caption{Principal Hugoniot of CH with initial state $\rho_0=$ 0.3 g/cm$^3$ and $Ti=$ 0.025 eV for SESAME equations of state used in our simulations to describe the aerogel.}\label{fig:hug}
\end{figure}

The ion temperature in the shocked region is around 10 eV as can be seen from the color plots Fig.\ref{fig:TIRO}. It goes rapidly to temperature around 100 eV in the ablated region. The shocked density remains between 1 and 1.5 g/cm$^3$ because, even if the temperature decreases as the shock proceeds through the aerogel (because it is not sustained), the density does not vary much because the thermodynamic state of the shocked matter just behind the shock lies on the principal Hugoniot with initial state at $\rho_0=$ 0.3 g/cm$^3$. As the temperature of shocked material decreases from 20 to 10 eV, the shocked density decreases much less, from 1.4 to 1.3 g/cm$^3$ (4.7 to 4.3 in compression), because of the proximity of the Hugoniot inversion (see Fig.\ref{fig:hug}).   

Radiographic images are simulated using the 3D monte-carlo particle transport code DIANE \cite{diane}. In this code, the result of the 2 dimensional FCI2 axisymetric radiation hydrodynamic simulation described earlier is made 3D by rotation around the axis. The following element have been simulated : a uniform source consisting of the simulated backlighter spectrum (at different times during the backlighter pulse), located at its actual experimental position, the pinhole (and its finite size) to image the tube, and filters on the line of sight to the CCD screen. 

\par 

Raw intensity profiles from the XRFC images cannot be compared easily to these simulations owing to the non uniform illumination of the shock-tube by the BL in the actual experiment. This complex illumination cannot be taken into account in the radiographic image simulations because it would require another diagnostic, which we had not, to visualize specifically that shape directly on the BL. The purpose of the next section is the removal of this detrimental background. This procedure is carried out on experimental intensity profiles, along the tube axis, of radiographic images.           

\begin{figure*}[t] 
\includegraphics[width=4cm]{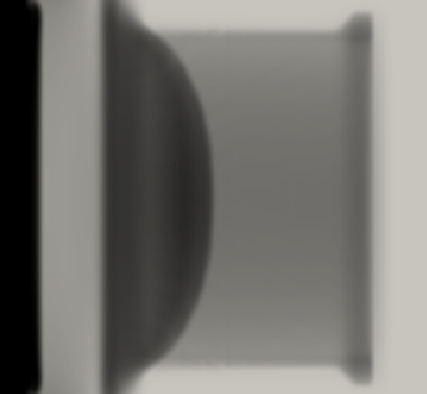}\hspace{.2 cm}\includegraphics[width=4cm]{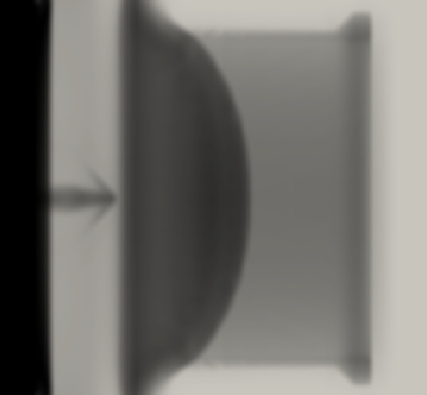}\hspace{.2 cm}\includegraphics[width=4cm]{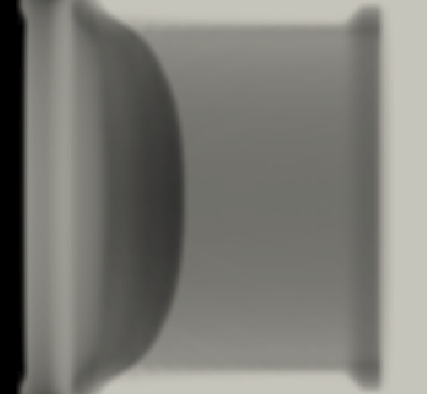}\hspace{.2 cm}\includegraphics[width=4cm]{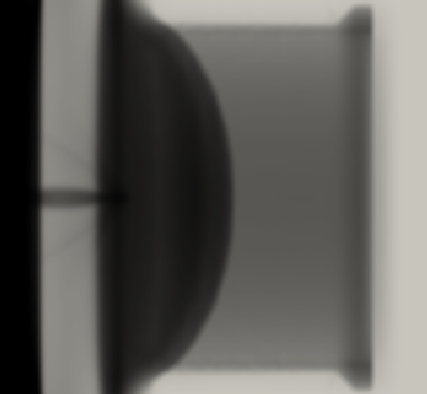}\\
\vspace{.1cm}
\includegraphics[width=4cm]{X1_PI_14ns.jpg}\hspace{.2 cm}\includegraphics[width=4cm]{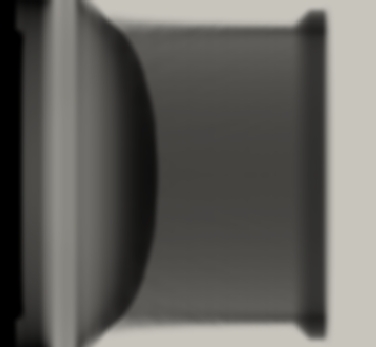}\hspace{.2 cm}\includegraphics[width=4cm]{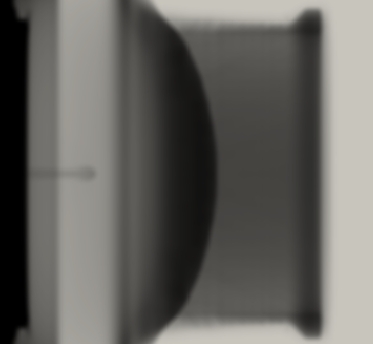}\hspace{.2 cm}\includegraphics[width=4cm]{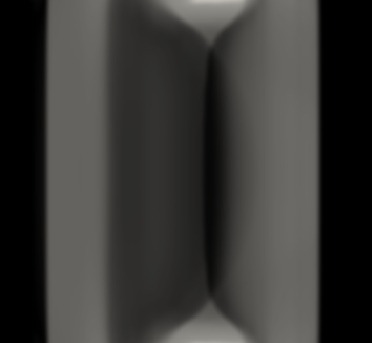}
\caption{DIANE simulations of the same shots and in the same order as in Fig.\ref{fig:rad}. (Top, from left to right) shots 70989, 70990, 70993 and 78804. (Bottom, from left to right) shots 78806, 78805, 78807 and 78808 (double-flux).}\label{fig:rad2}
\end{figure*}

\section{\label{sec:xrfc} Experimental results}

Precise results are obtained with the XRFC diagnostic. The discussion of the SOP results are deferred to section \ref{sec:sop}. 

\par 

The X-ray framing camera (XRFC) is used to obtain 2D side-on radiographic images of the tube. It was configured with a four strip microchannel plate (MCP). It converts x-rays into photoelectrons with a gold photocathod coated on its surface. The photoelectrons are then accelerated through the microchannels before they hit a phosphor plate creating visible light that is recorded on a 4096$\times$4096 CCD screen. On each strip, four images are separated by $\sim$50 ps and each strip is delayed by 200 ps. It results in 16 images recorded on the CCD screen (see Fig.\ref{fig:mcp} for an example) ranging from 0 to 750 ps from the beginning of the backlighter pulse (to be precise, $t$=0 of the backlighter pulse corresponds in fact to the time when the laser pulse reaches 3$\%$ of its predicted peak intensity).

\par

On Fig.\ref{fig:rad}, a selection of radiographic images is represented. On images corresponding to {\it simple flux} with piston ($d_\mathrm{ab}>0$), the shocked region (in dark in the middle) is separated from the unshocked region on the right by a curved shock and it is separated from the ablated region on the left by a straight ablation front. The curvature of the shock, also mentioned in \cite{lstRT}, was also found in our simulations Fig.\ref{fig:rad2} (as mentioned in section \ref{sec:hydro}). The straight form of the ablation front may come as a surprise. Indeed, x-rays come from a hole smaller than the section of the tube and so the distribution of x-rays should be inhomogeneous on the front. It is indeed the case but, because the ablated velocity is large, any perturbation on the front is rapidly damped. The agreement between simulations, Fig.\ref{fig:rad2}, and experiments, Fig.\ref{fig:rad}, is qualitatively good but in the next subsections we would like to make this comparison more quantitative. 

\par 

The intensity profiles are very good candidates for they give access to important information, such as density contrast and precise position of shocks and ablation fronts for instance. Together with SOP results (probing velocities of shocks, see section \ref{sec:sop}), this can allow to get few points of the equation of state of the material passed through by shocks, namely, the aerogel. Getting these informations is the purpose of the next few sections.

\subsection{Selection of images for background removal}

Experimental radiographic images are marred by an important background noise originating from the backlighter. The inhomogeneity and coherence of the laser light hitting the backlighter imprints a field of micrometer-sized hot-spots that shows of on the radiographic images. In addition to these small scale inhomogeneities superimposes a large scale inhomogeneity ($\sim$ 800-900 $\mu$m) due to the finiteness of the laser spot on the backlighter. In order to compare the simulated radiography (assuming a homogeneous radiation from the backlighter) with the experimental radiography it is essential to remove the large scale inhomogeneity from the experimental radiography.

\par 

In order to remove that background, an hypothesis on its general shape needs to be agreed on. It is customary to assume that the intensity distribution of a laser across its section is a supergaussian. When backlighted through a high Z material, such as titanium, vanadium, scandium or iron, it produces a distribution of x-rays which is also a supergaussian to a very good approximation. Of course, the supergaussian so produced does not have the same parameters. But, to proceed further into the analysis, the only thing to bear in mind is that the shape of the incident intensity profile is a supergaussian with some parameters to be adjusted:
\begin{align}
f(z, I_0, z_c, \sigma, n)=I_0\,\exp\left[-\left(\frac{\mid z-z_c\mid}{\sigma}\right)^n\right]\label{ffit}
\end{align} where $I_0$ is the intensity transmitted at the center of the laser spot, $n$ is the power of the supergaussian, $\sigma$ is its half width (the radius of the laser spot) and $z_c$ the location of its center.

\par

\begin{figure*}[t] 
\includegraphics[width=4cm]{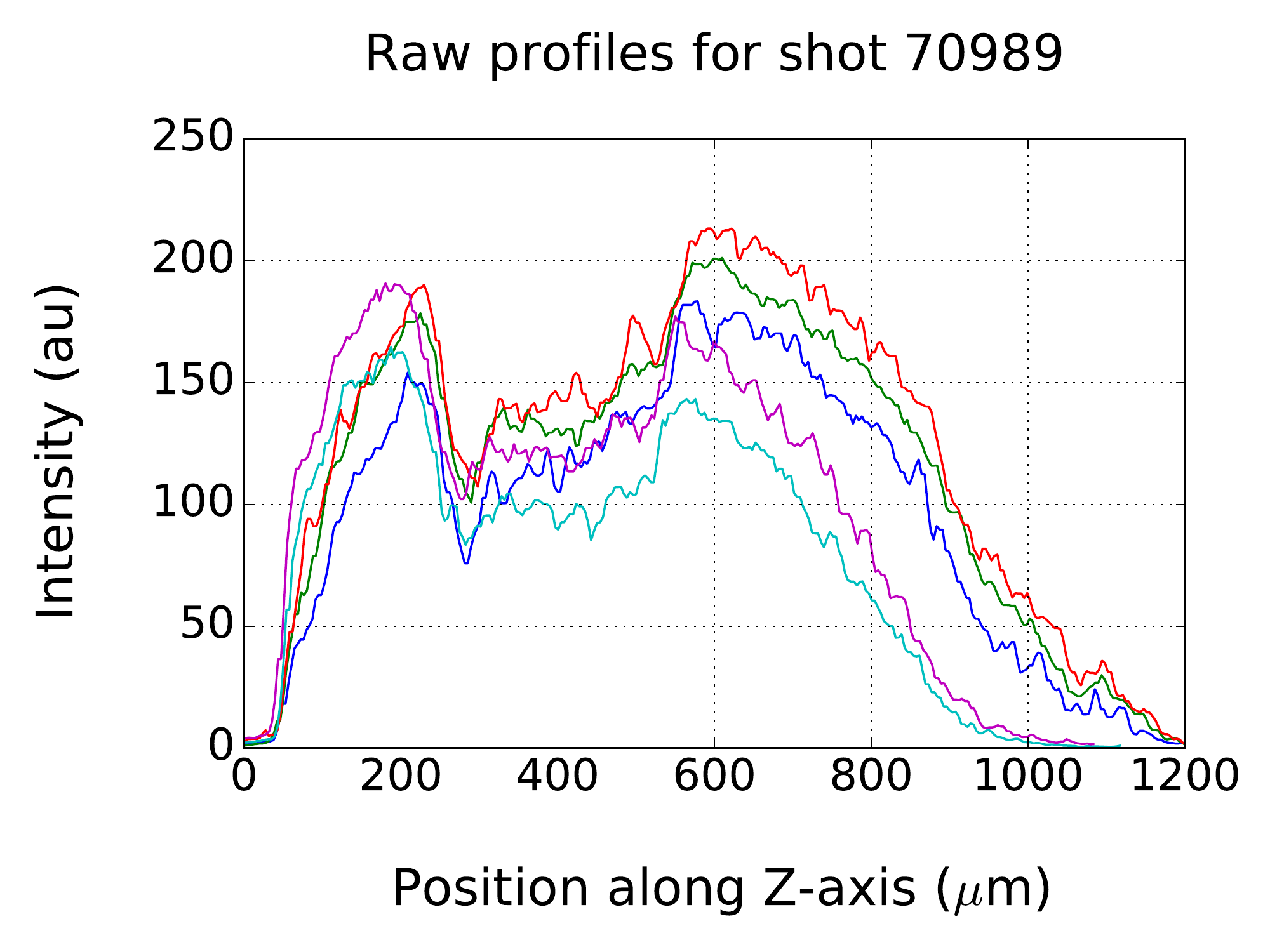}\hspace{.2 cm}\includegraphics[width=4cm]{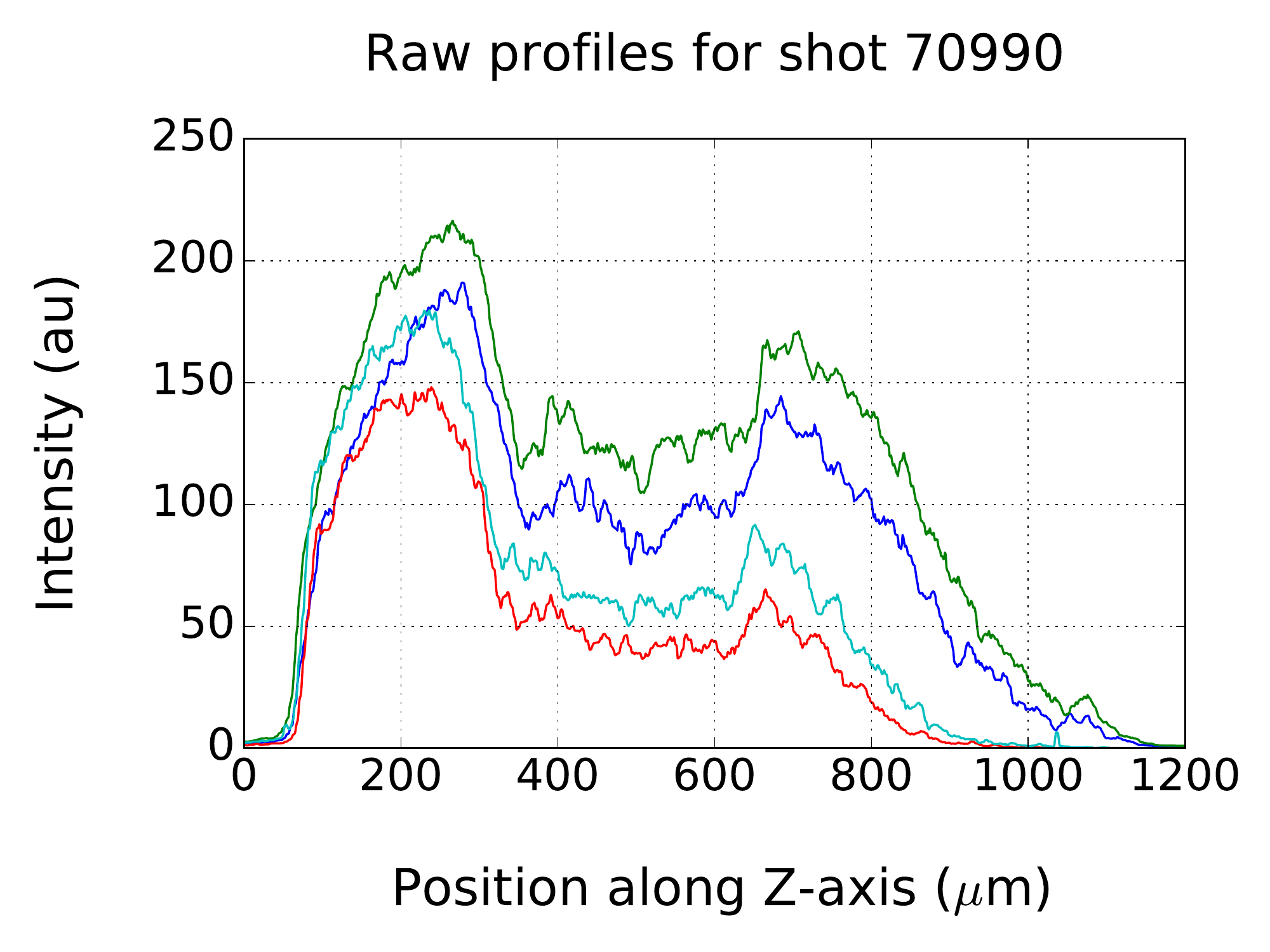}\hspace{.2 cm}\includegraphics[width=4cm]{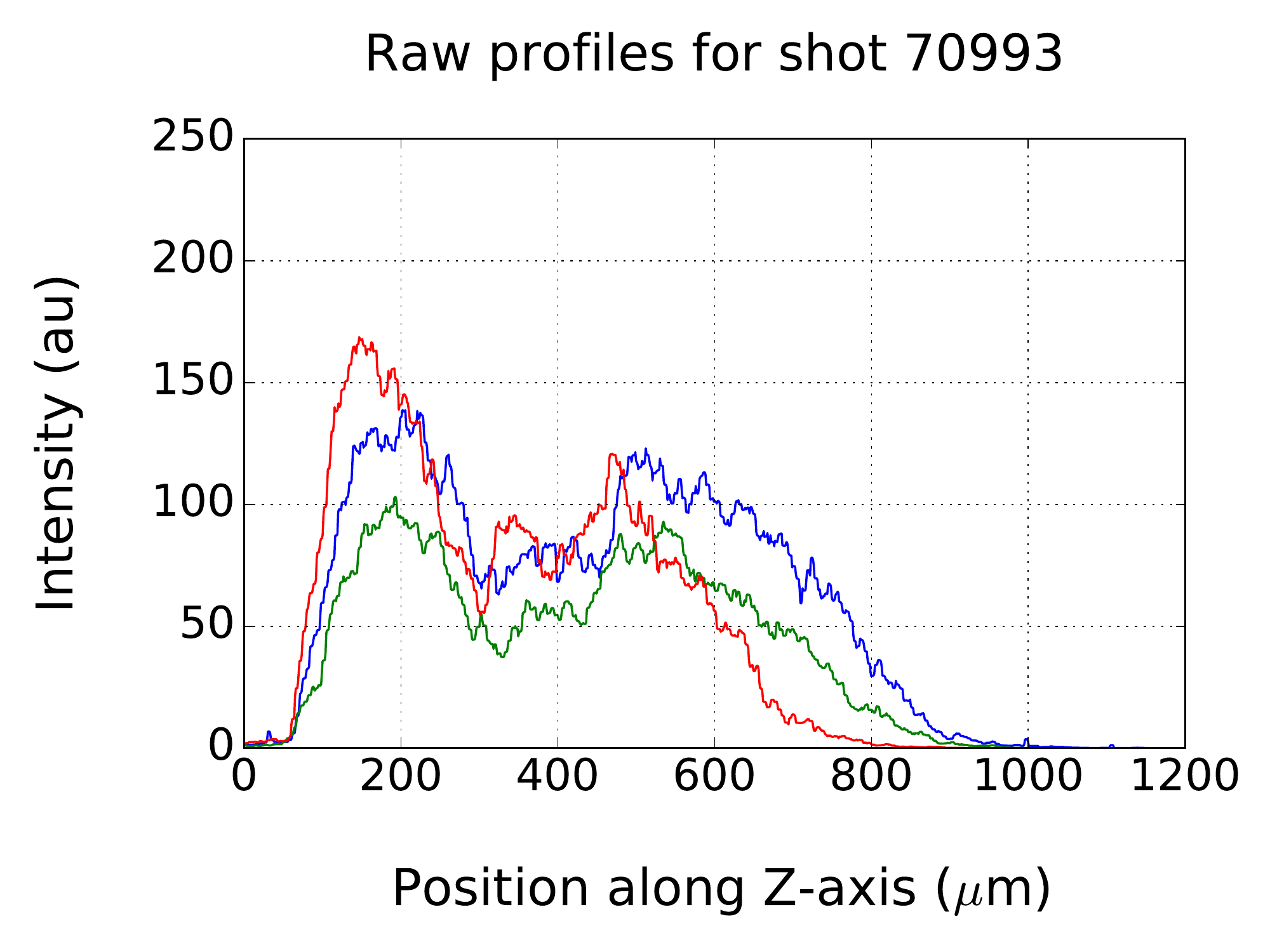}\hspace{.2 cm}\includegraphics[width=4cm]{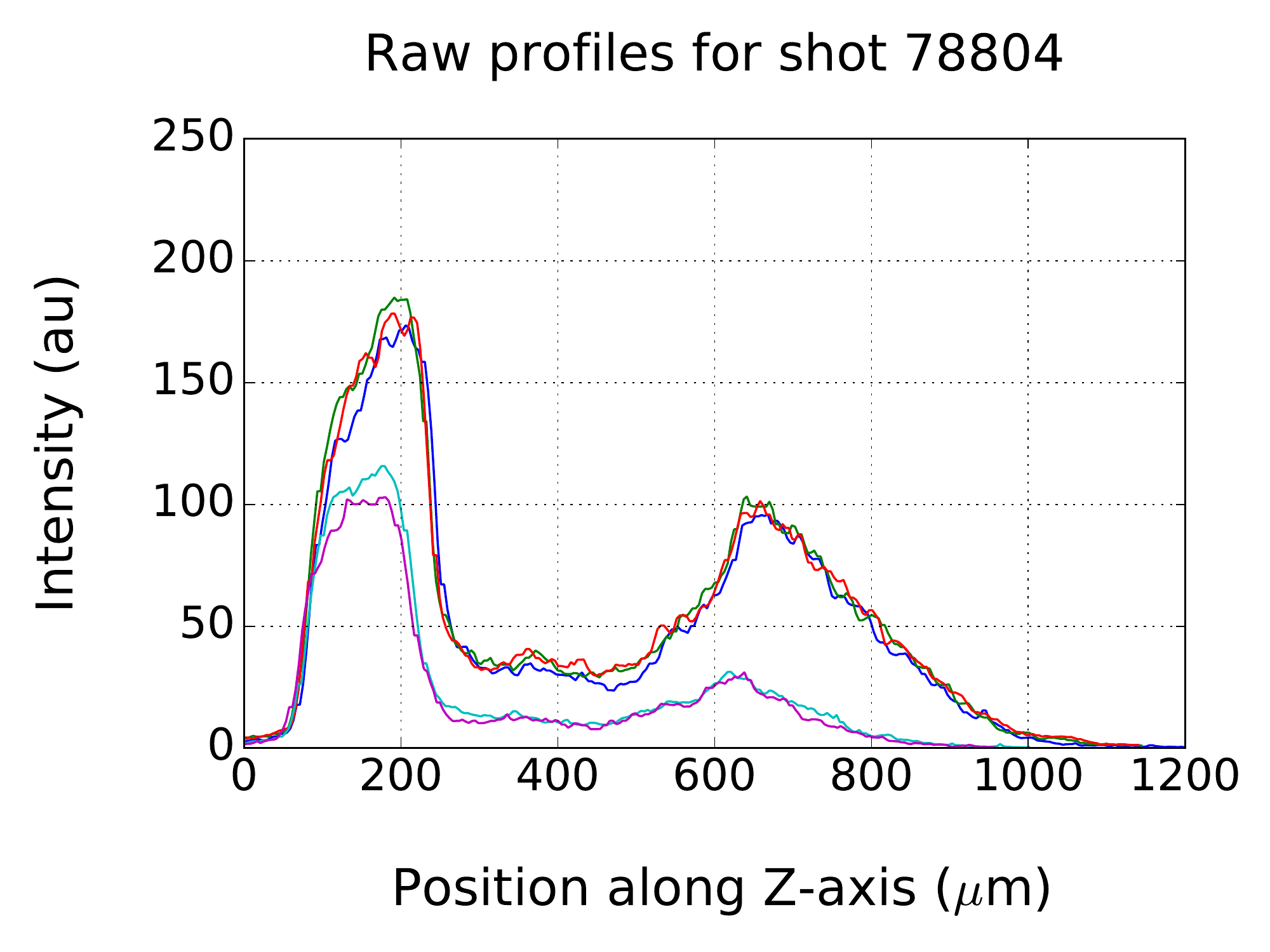}\\
\vspace{.1cm}
\includegraphics[width=4cm]{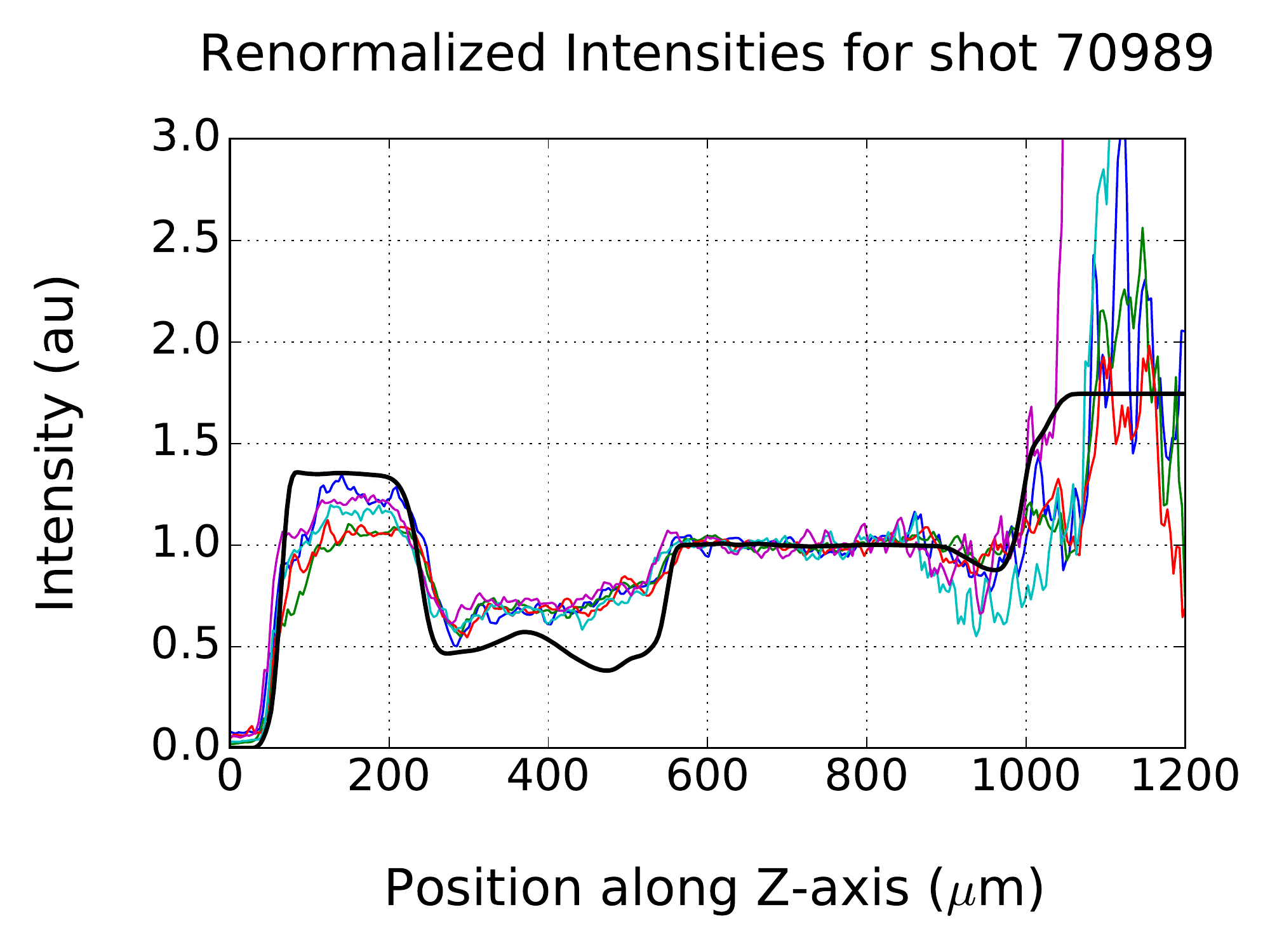}\hspace{.2 cm}\includegraphics[width=4cm]{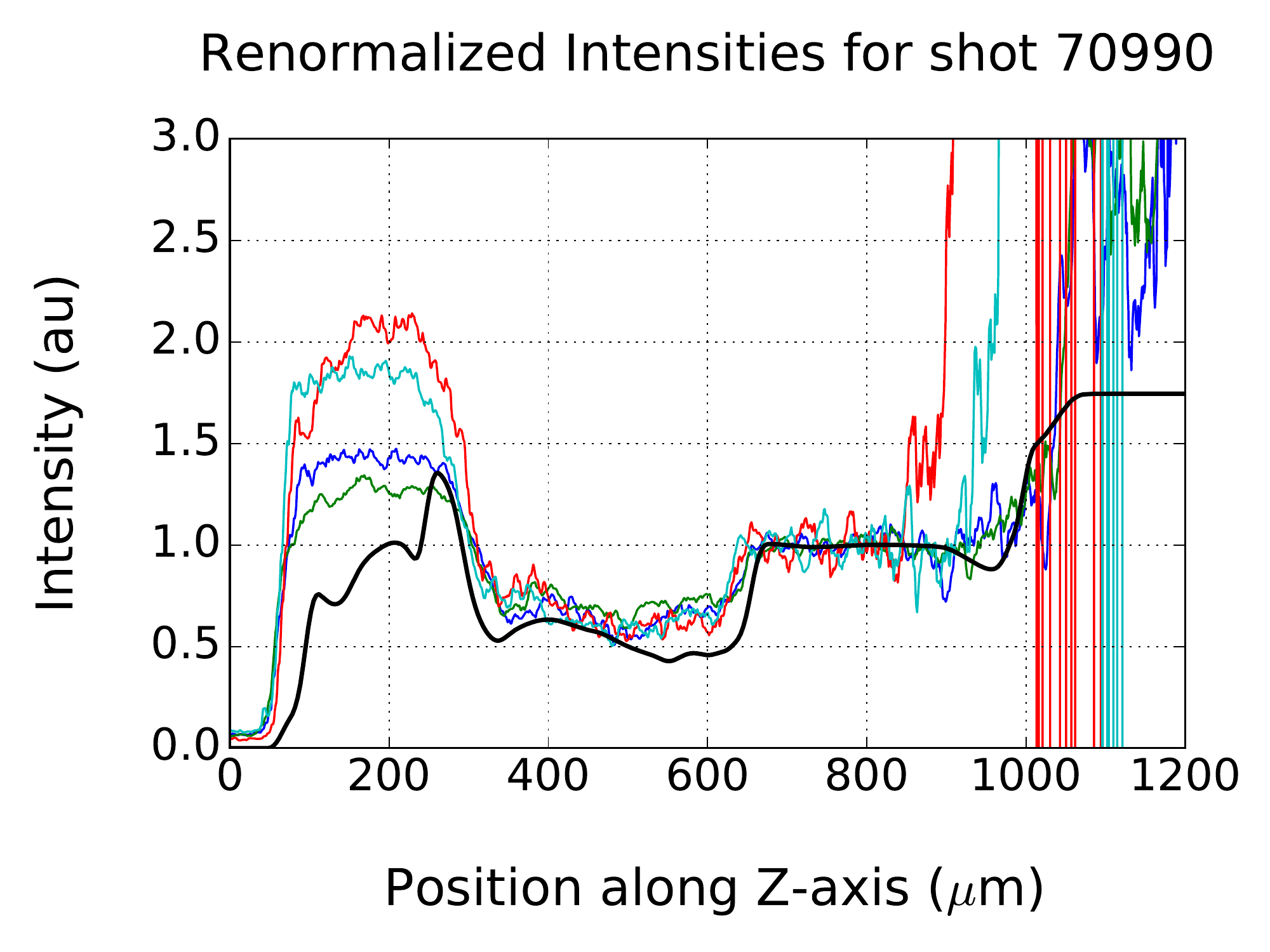}\hspace{.2 cm}\includegraphics[width=4cm]{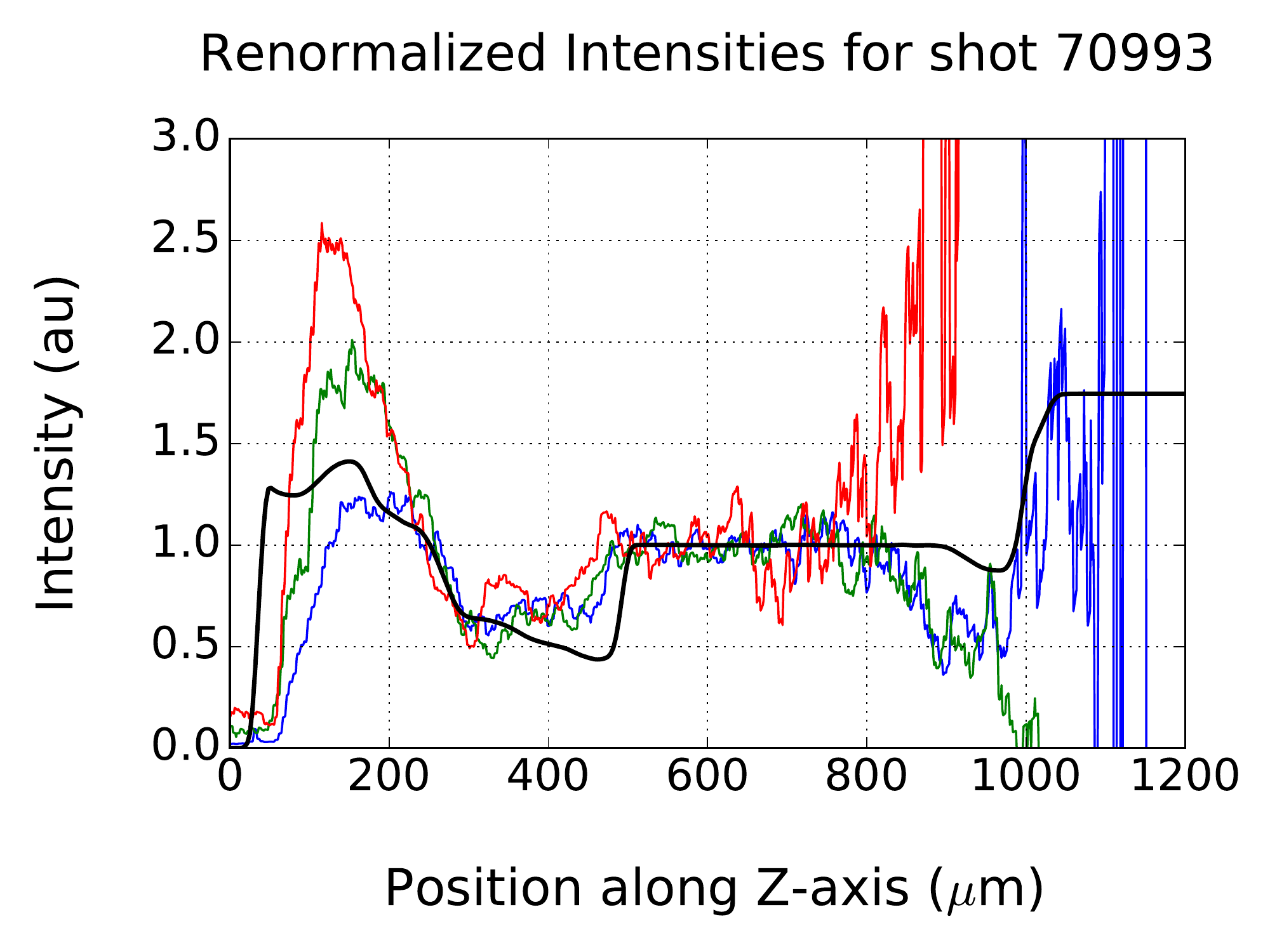}\hspace{.2 cm}\includegraphics[width=4cm]{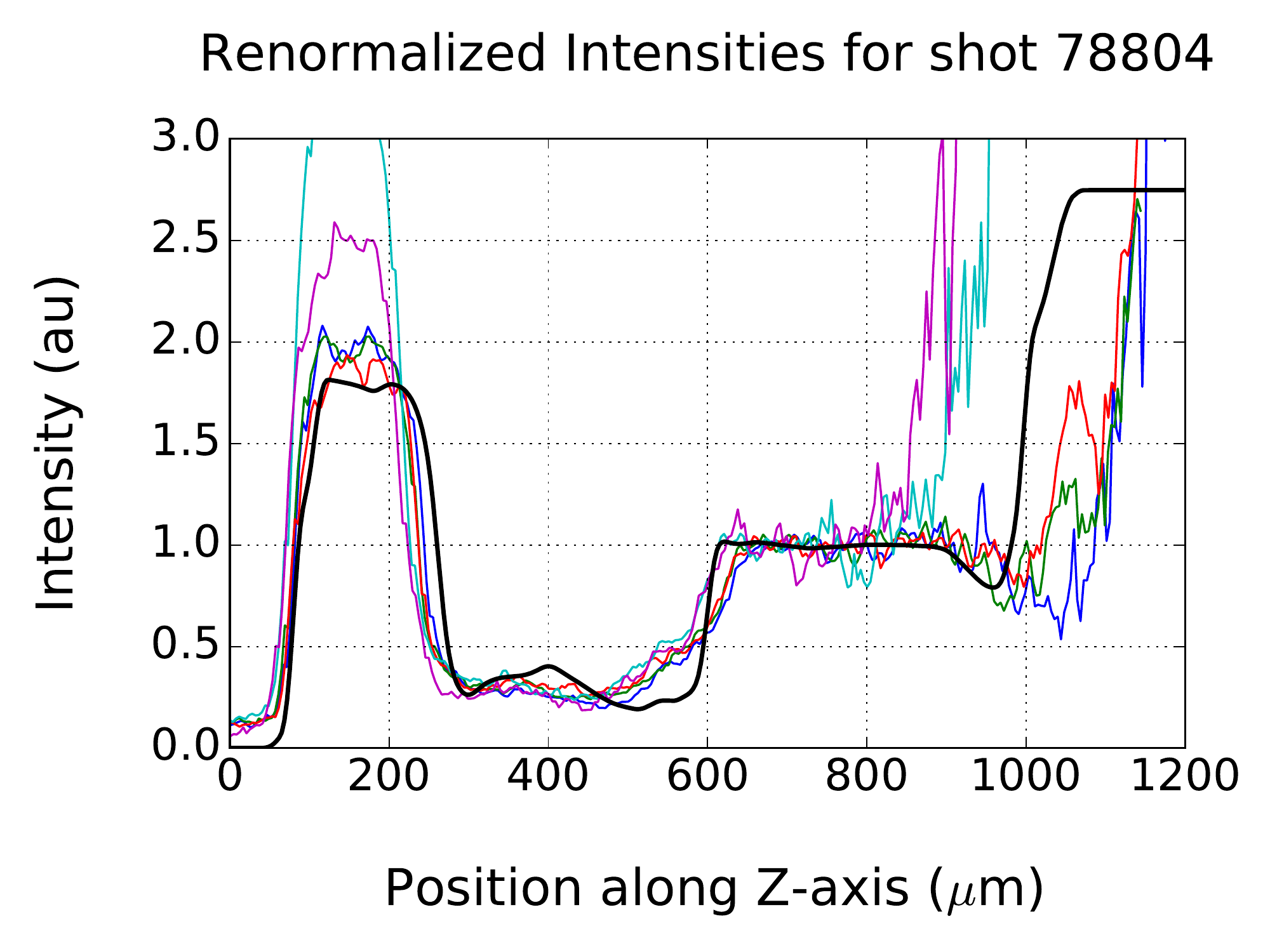}
\vspace{.1cm}
\includegraphics[width=4cm]{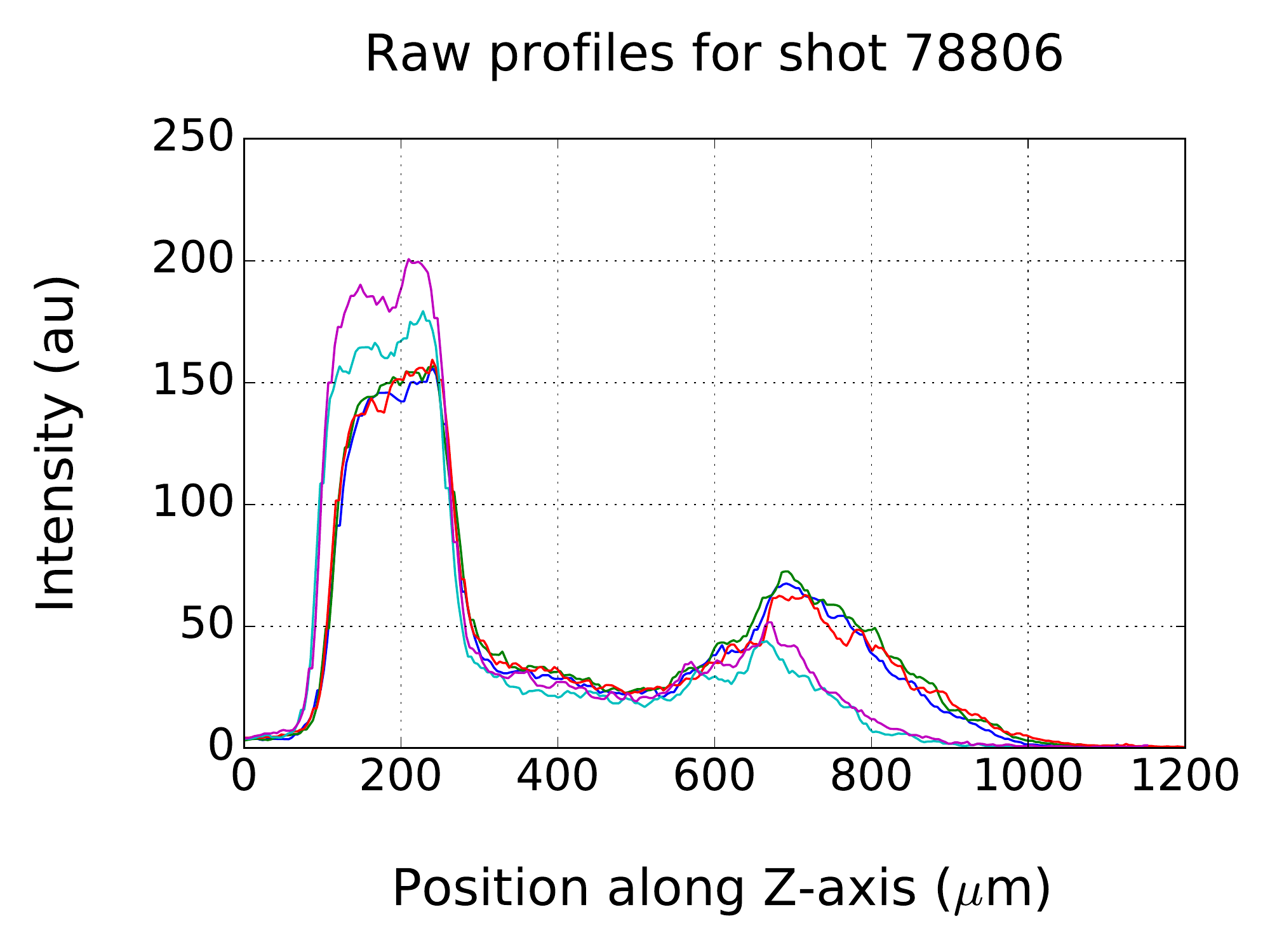}\hspace{.2 cm}\includegraphics[width=4cm]{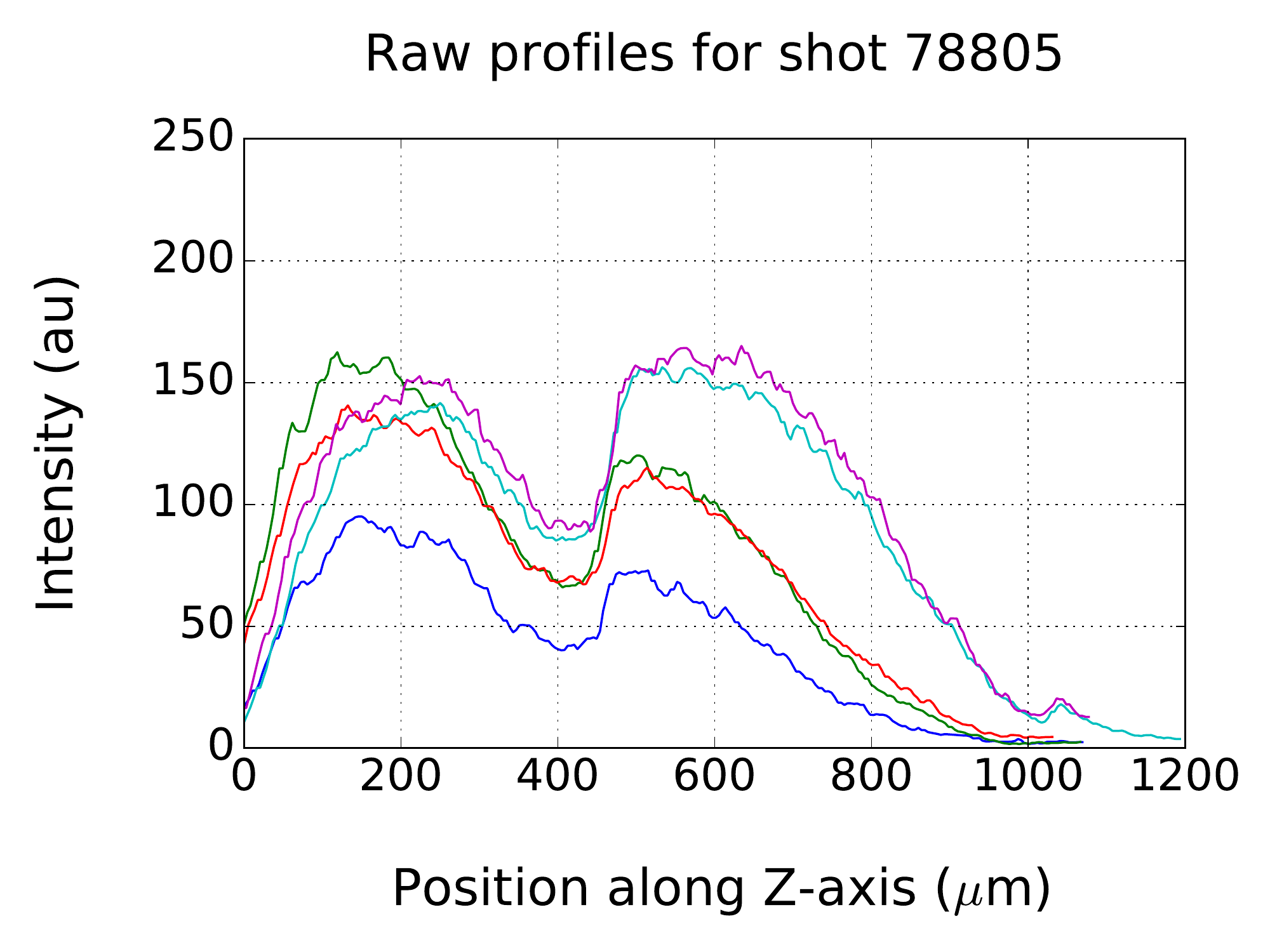}\hspace{.2 cm}\includegraphics[width=4cm]{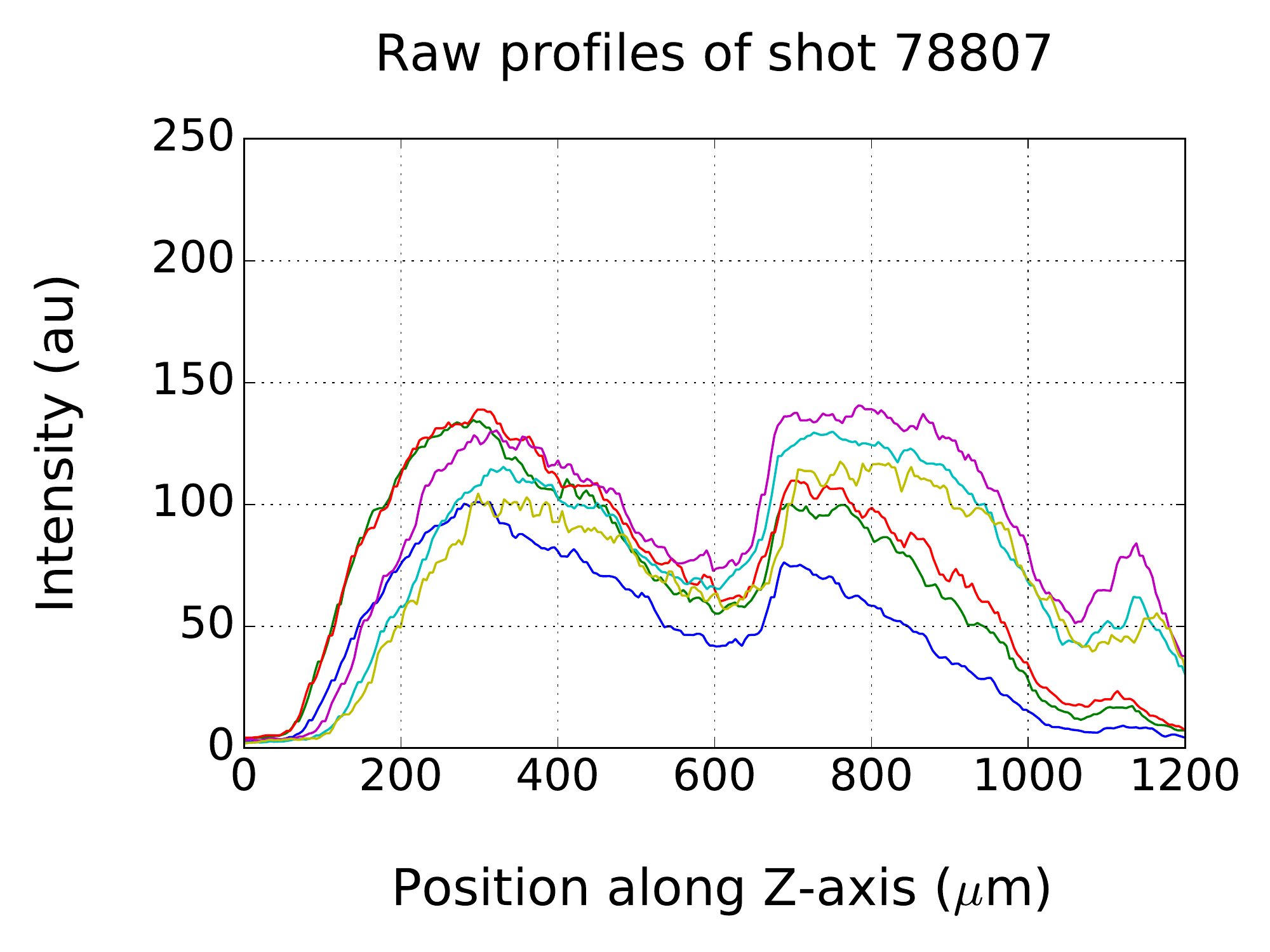}\hspace{.2 cm}\includegraphics[width=4cm]{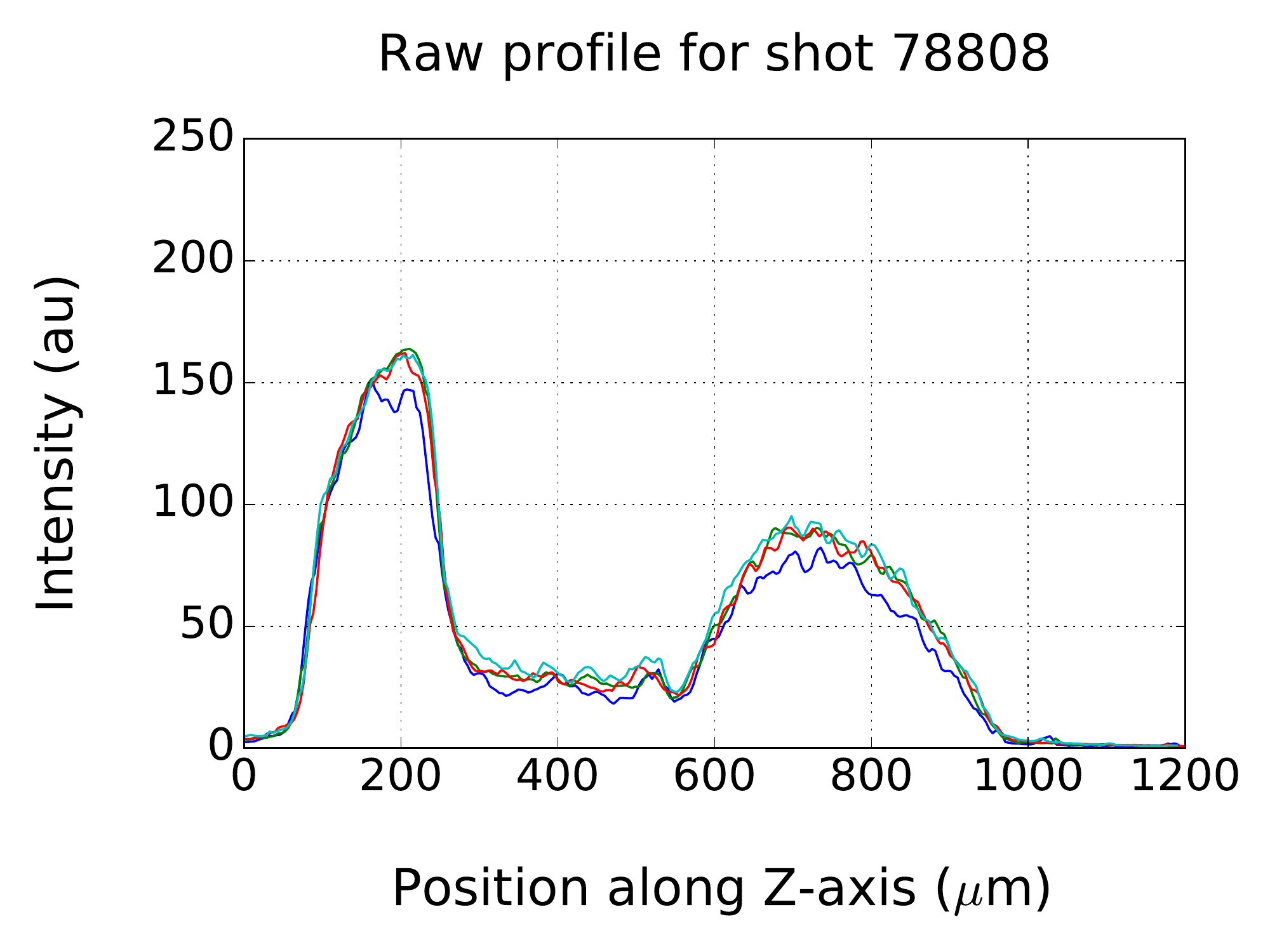}\\
\vspace{.1cm}
\includegraphics[width=4cm]{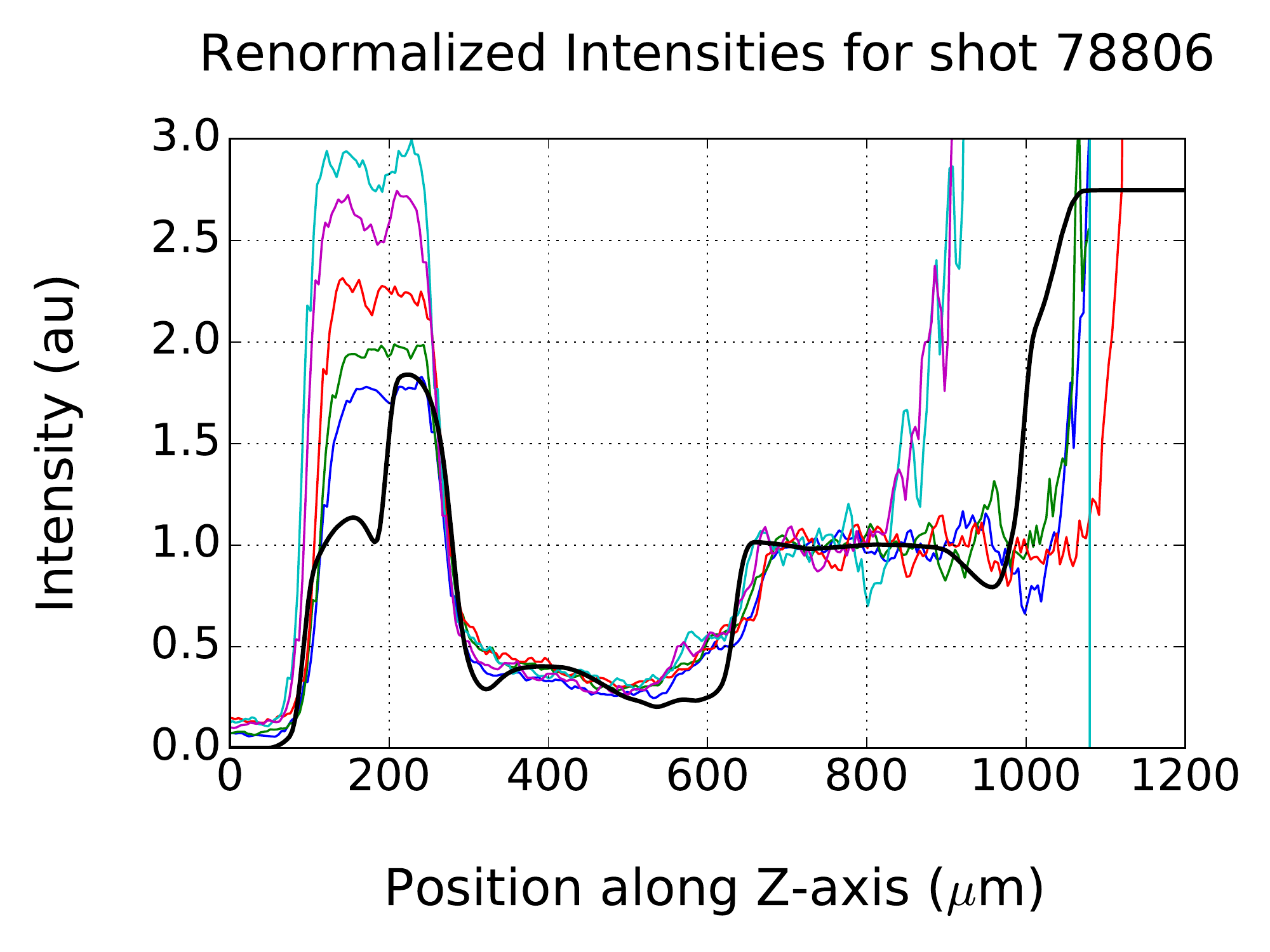}\hspace{.2 cm}\includegraphics[width=4cm]{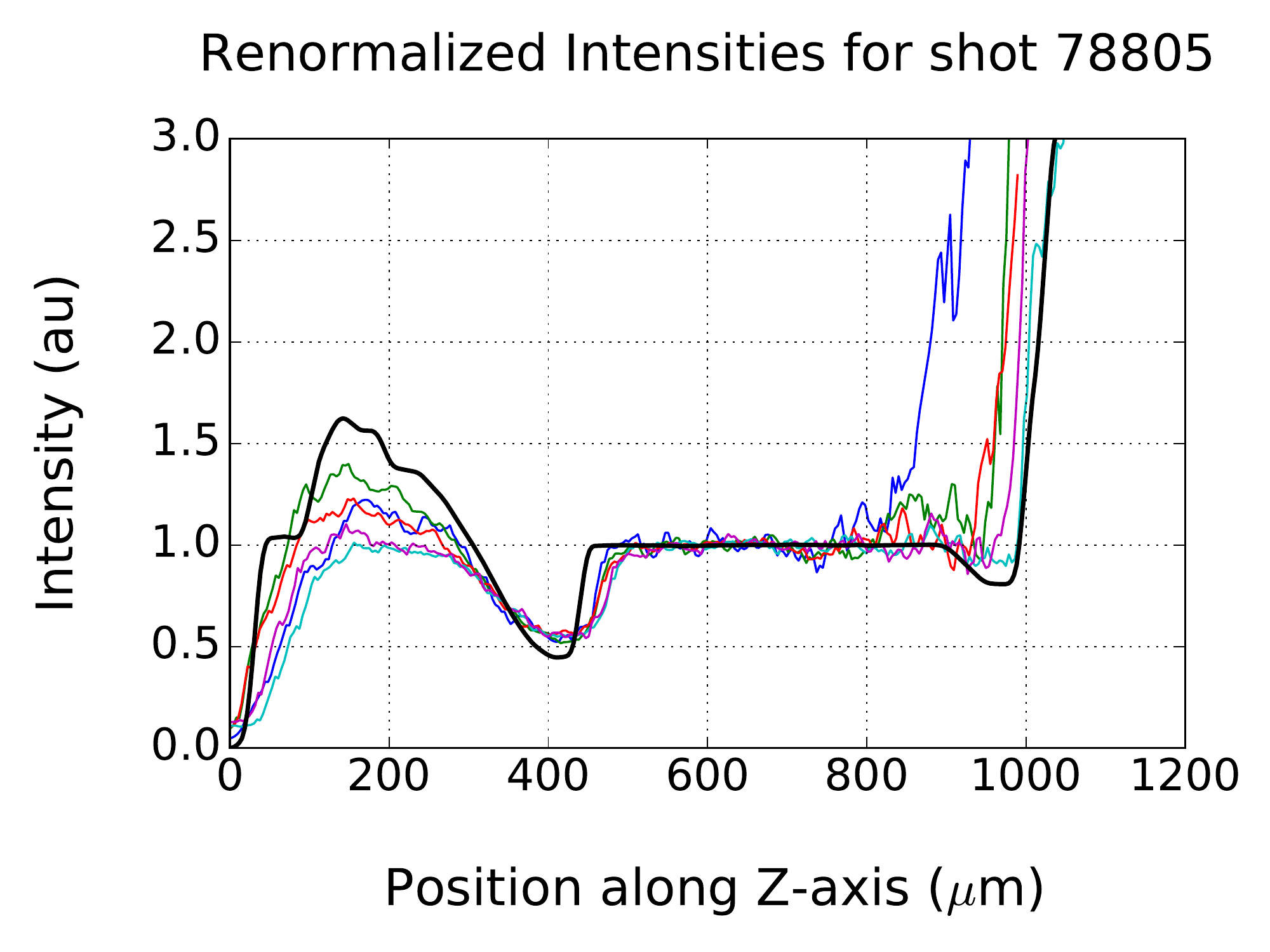}\hspace{.2 cm}\includegraphics[width=4cm]{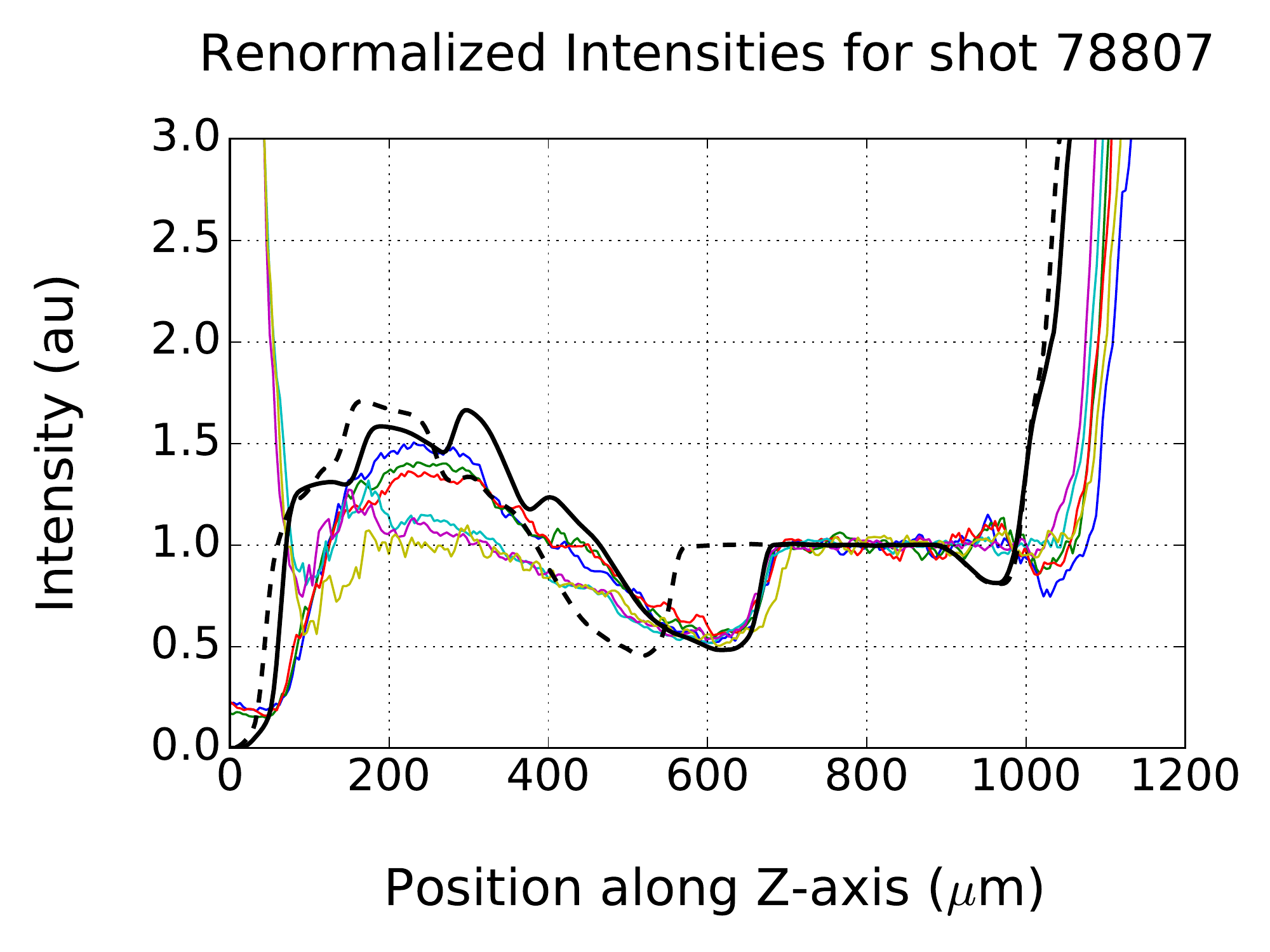}\hspace{.2 cm}\includegraphics[width=4cm]{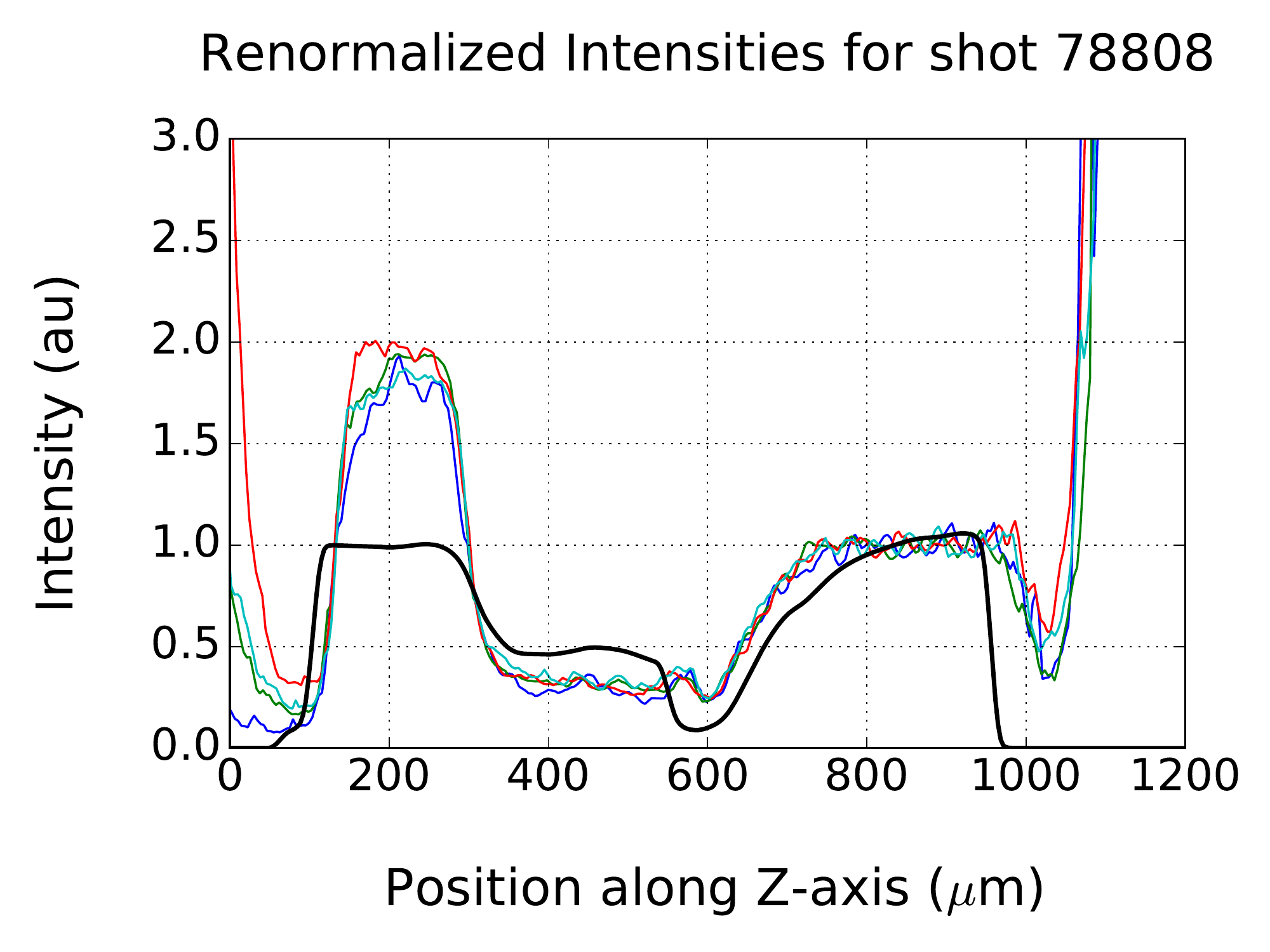}\\
\caption{Raw intensity profiles extracted from different shots and the associated renormalized intensity underneath. Superimposed to each bundle of renormalized intensity profiles is a dark curve representing the DIANE post-shot simulation of the renormalized intensity profile (assuming the source is spatially uniform) with the corresponding, $K_\alpha$ only, backlighter spectrum. The dashed black line in the graph of the renormalized intensity profile of shot 78807 corresponds to the simulation at the expected time of 7 ns whereas the solid black line corresponds to the simulation at $t\approx$ 9 ns that is the best match to the experimental data at 7 ns. The discrepancy is due to the SESAME equation of state that is certainly inappropriate for our aerogel (see end of section \ref{sec:sop}).}\label{fig:radprof}
\end{figure*}

In order to get a precise reconstruction of the background, it has empirically been found that each selected images for the background removal should comply with a few reasonable constraints: 
\begin{itemize}
\setlength{\itemsep}{-0.1cm}
 \item[(i)] the fiducial grid must be visible (in order to adjust position and scale of the profile),
 \item[(ii)] the axis of the tube should be visible and defect-less,
 \item[(iii)] the signal to noise ratio (SNR) should be at a reasonable level (SNR $\gtrsim$ 10).
\end{itemize} These constraints have enable to select on average 4 to 6 images out of the 16 of each shot as can be checked on Tab.\ref{tab:table1} ($N_\mathrm{img}$ is the number of images that went through this analysis process). These constraints are also the reason why none of the iron backlighted images have been selected because the SNR condition was poor. As an example of a lower limit SNR on intensity profiles extracted from radiographic images, one can compare the raw profiles of shot 70993 (SNR on the edge of what is acceptable) with others (reasonable to good SNR) on Fig.\ref{fig:radprof}.

\subsection{\label{sec:remov} Background removal method}

\begin{figure}[h] 
\includegraphics[width=8cm]{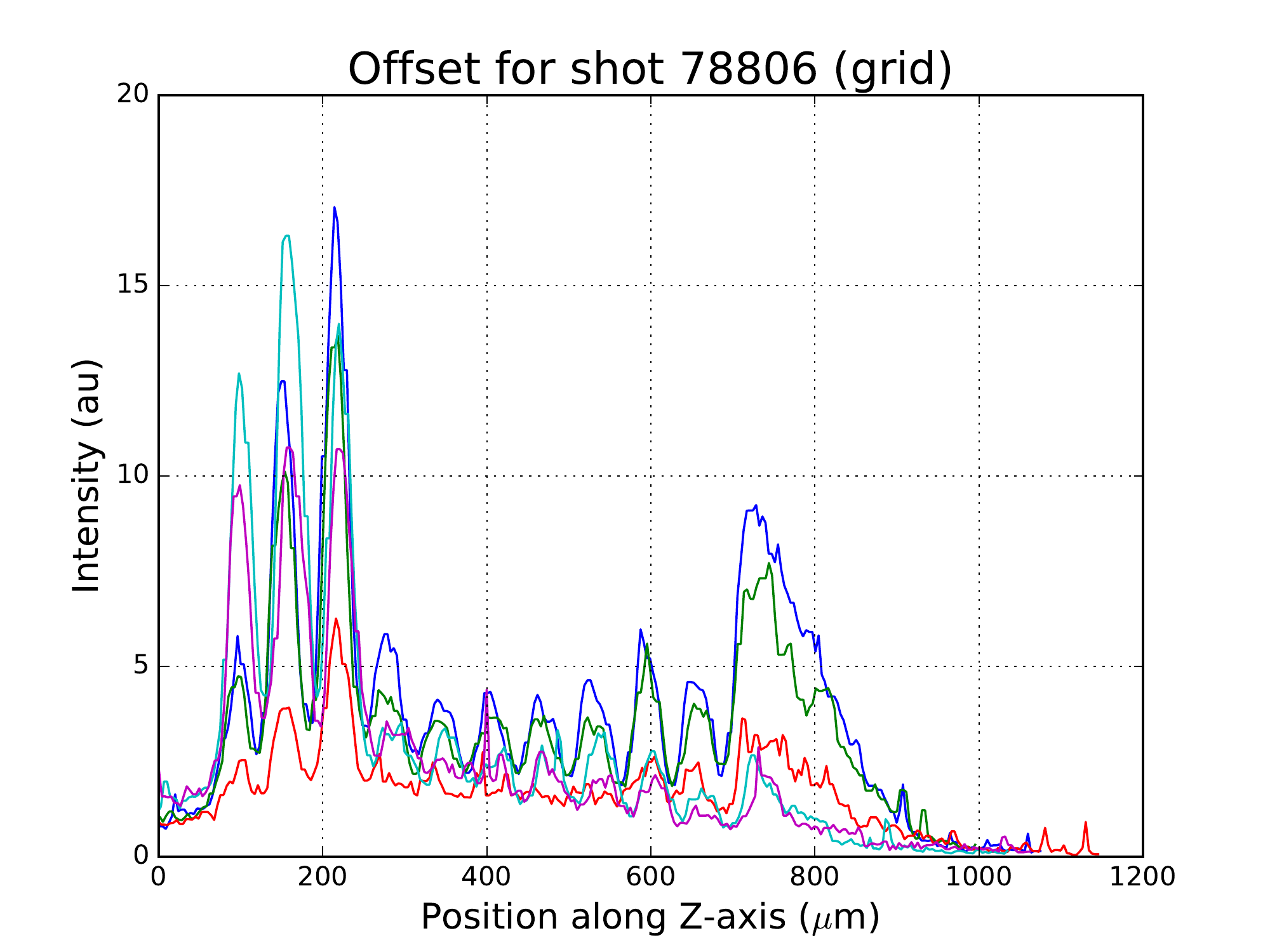}
\caption{(Color online) Profiles of the grid for the five different radiographic images of shot 78806 (as an example) once at scale and shifted to the actual experimental position along the z-axis of the shock tube.}\label{fig:grd1}
\end{figure}

Let us go through all the steps of the analysis on the example of shot 78806. 

\par

The first step consists in selecting all images complying with the constraints enacted earlier. It represents, for this particular shot, 5 images out of the 16 images on the CCD screen of the XRFC. On all 5 selected images, the fiducial grid is visible and an intensity profile is plotted for each grid of each image. Since we know the spacing between each square of the grid (62 $\mu$m) and that there are 11 squares, for 2015 grids (16 squares for 2013 grids), along the axis of the experiment and that the first square is in contact with the halfraum gold wall (in all but shot 70993), a precise resizing and repositioning of each profile is performed to fit to the real units (cf. Fig.\ref{fig:grd1}). A spike is missing between 0 and 100 $\mu$m because the first square in contact with the halfraum is masked by some of its ablated gold at that time.

\begin{figure}[h] 
\includegraphics[width=8cm]{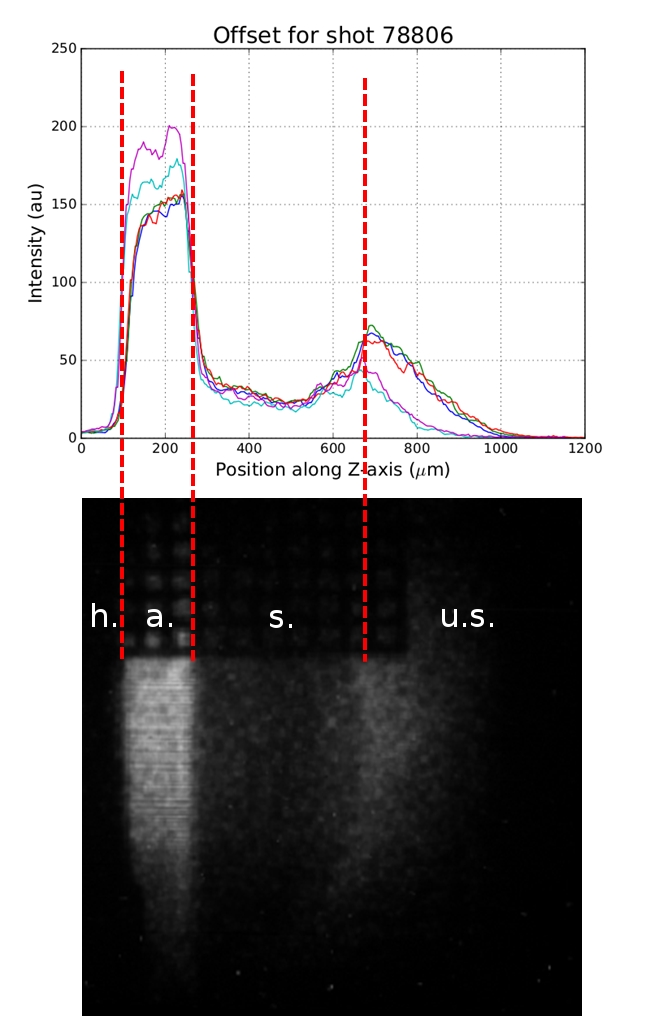}
\caption{(Color online) Intensity profiles of the five different radiographic images scaled and shifted by the same amount as the grid profiles.}\label{fig:grd2}
\end{figure}

\par 

In the second step, the raw intensity profiles along the axis are plotted (Fig.\ref{fig:grd2}) in the real coordinate basis aforementioned. These intensity profiles can be divided into four parts: (h.) represents the halfraum or some ablated gold from the halfraum that is opaque. The part designated by (a.) is the ablated plastic material, from the piston and aerogel, which is the more transparent part. The shocked material is designated by (s.). The unshocked material, which is always in the aerogel, is designated by (u.s.) and will be the reference state for what follows. 

\par

The third step has to do with the function (\ref{ffit}) to be fitted. Clearly, (h.) and (s.) cannot be part of the fit. The first one because it blocks the radiation from the backlighter, it is then uncorrelated to (\ref{ffit}), and the second one because there is no {\it a priori} knowledge about the thermodynamical state of that region (It is precisely what needs to be inferred from the transmission). At the opposite, points in (u.s.) should be part of the fit since we know everything about their state: the density and opacity are uniform, and still have their initial value (since unaffected by the hydrodynamic and barely affected by preheat as shown by simulations), and so is the transmission. Let us call this collection of points in the unshocked material $\left\{(z_\alpha, I_\alpha)\right\}_{\alpha\in [1,N_\mathrm{us}]}$ where $N_\mathrm{us}$ is the number of points selected in the unshocked zone and where $z_\alpha$ and $I_\alpha$ are respectively the location along the z axis and the experimental intensity in arbitrary unit at the point labeled $\alpha$. The points in (a.) must also be part of the fit but need to be corrected by a factor (unknown at this stage), called $\eta$ and to be determined later on, that takes into account the fact that the transmissions in (a.) and (u.s.) are not the same. This points selected in (a.) form a collection as well, designated by $\left\{(z_\beta, I_\beta)\right\}_{\alpha\in [1,N_\mathrm{ab}]}$ where $N_\mathrm{ab}$ is the number of points selected in the ablated zone (a.), and $z_\beta$ and $I_\beta$ are respectively the location along the z axis and the experimental intensity in arbitrary unit at points labeled $\beta$. 

\par

\begin{figure}[h] 
\includegraphics[width=8cm]{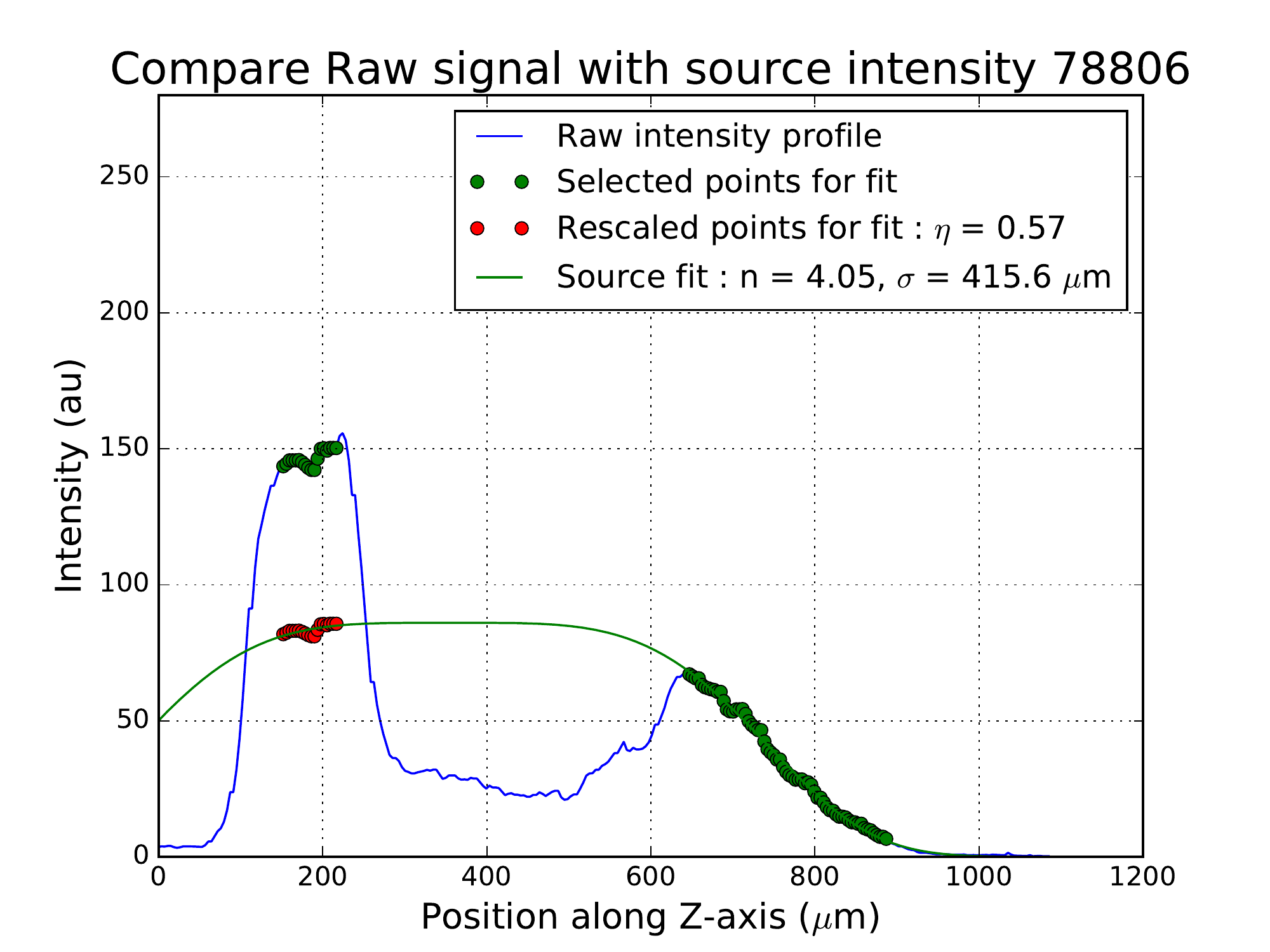} 
\includegraphics[width=8cm]{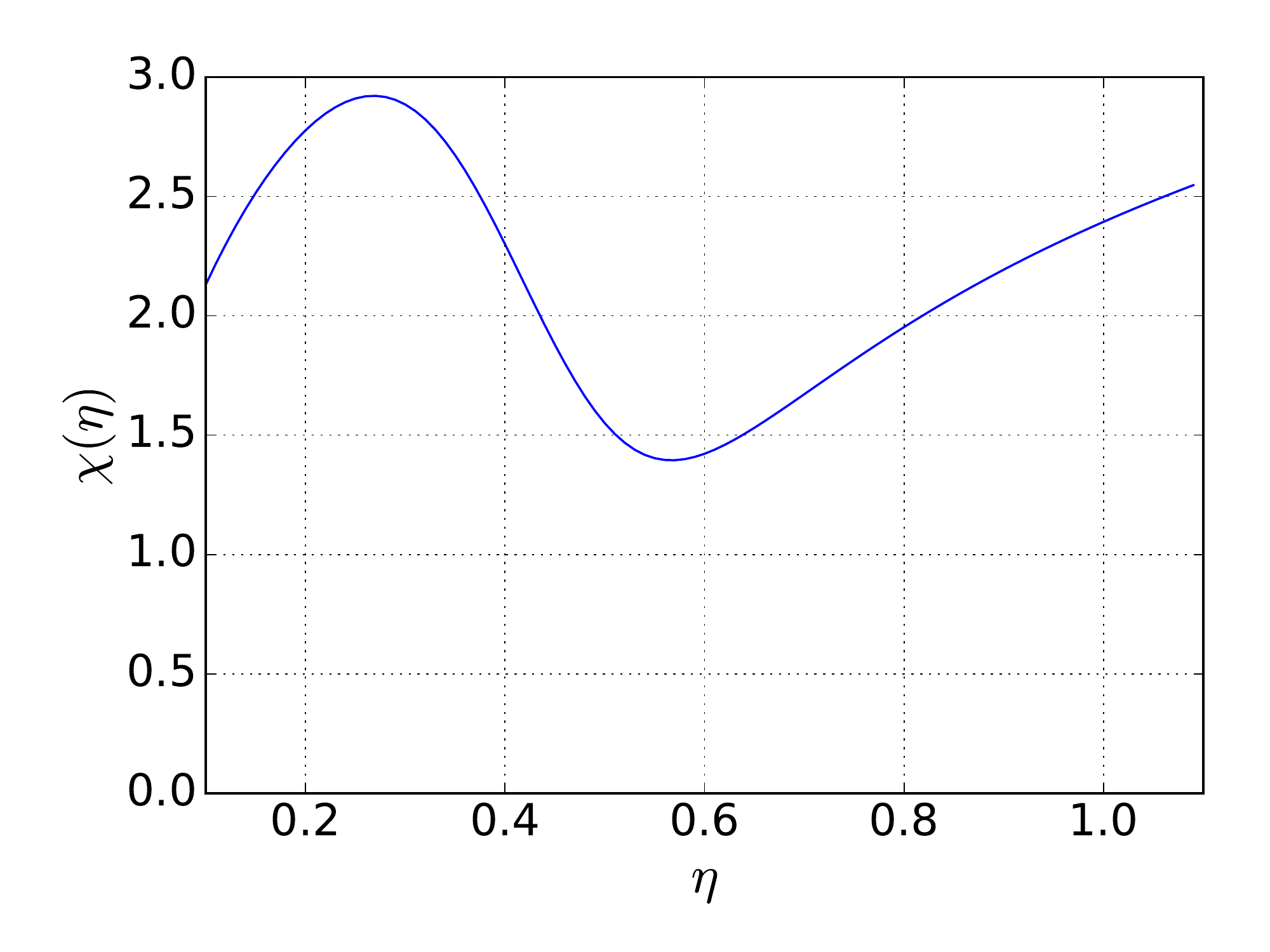}
\caption{(Color online) The top figure represents the various ingredients required to infer the backlighter profile : the blue curve is the raw experimental profile of intensity along the z-axis of the shock tube for one of the 5 radiographic images selected on shot 78806, the green points are the selected points for best fit in the ablated zone (on the left) and in the unshocked zone (on the right), the red points are the corrected selected points in the ablated region an the green curve is the best fit with a supergaussian.
The bottom figure is a typical representation of $\chi(\eta)$ with a minimum at the best value for $\eta$ ($\eta=$ 0.57 in this particular example).}\label{fig:prof1}
\end{figure}

The procedure to determine $\eta$ constitutes the fourth step. Every possible values of $\eta$ is tested (between 0.1 and 1.5 in practice). The best fit of $f$ is evaluated over the $N=N_\mathrm{us}+N_\mathrm{ab}$ experimental points selected in (u.s.) and corrected in (a.), namely 
\begin{align}
 \left\{(z_\alpha, I_\alpha)\right\}_{\alpha\in [1,N_\mathrm{us}]}\,\bigcup\,\left\{(z_\beta, \eta\,I_\beta)\right\}_{\beta\in [1,N_\mathrm{ab}]},
\end{align} notice the coefficient $\eta$ in front of $I_\beta$. This best fit (using a least square method) produces, for each value of $\eta$, an optimal value for $I_0$, $z_c$, $\sigma$ and $n$ in function (\ref{ffit}). Obviously, these values depend upon $\eta$. Now, among all the tested values of $\eta$, the one that is optimum (the closest match to the corrected data) is determined by finding a local minimum to the mean squared deviation between the corrected data and the fit
\begin{align}
\chi^2(\eta)&=\sum_{\alpha\in (\mathrm{us})} \frac{\left(I_\alpha-f_\alpha\right)^2}{N}+\sum_{\beta\in (\mathrm{ab})} \frac{\left(\eta\, I_\beta-f_\beta\right)^2}{N},\label{chi}
\end{align} where
\begin{align}
f_\alpha&=f(z_\alpha, I_0(\eta), z_c(\eta), \sigma(\eta), n(\eta)),\\
f_\beta&=f(z_\beta, I_0(\eta), z_c(\eta), \sigma(\eta), n(\eta)).
\end{align} Each point used for the fit has the same weight as the others in the cost function $\chi^2$ in eq.(\ref{chi} ) regardless of its origin (from (u.s.) or from (a.)). An example, for shot 78806, is given in Fig.\ref{fig:prof1}. It is comforting to know that, for every profile of any selected images, only one local minimum was generated in function $\chi(\eta)$, such as the one depicted in Fig.\ref{fig:prof1}. In this figure, if the curve was extended to values of $\eta<0.1$ a global minimum would have been found. It is an uninteresting result that corresponds to a bell shape (small value of $\sigma$) that goes through the (u.s.) points and bents down rapidly to return to zero for (a.) corrected points. By sticking with the local minimum, almost every value of $\sigma$ we have found is close to 400 $\mu$m which corresponds to the half diameter of the laser spot on the BL. Differences are due to the fact that, in general, the intersection of the shock-tube axis with the BL circular (or elliptical) spot, as seen from the radiographic line of sight, is not a diameter but a chord with a smaller length.

\begin{figure}[h] 
\includegraphics[width=8cm]{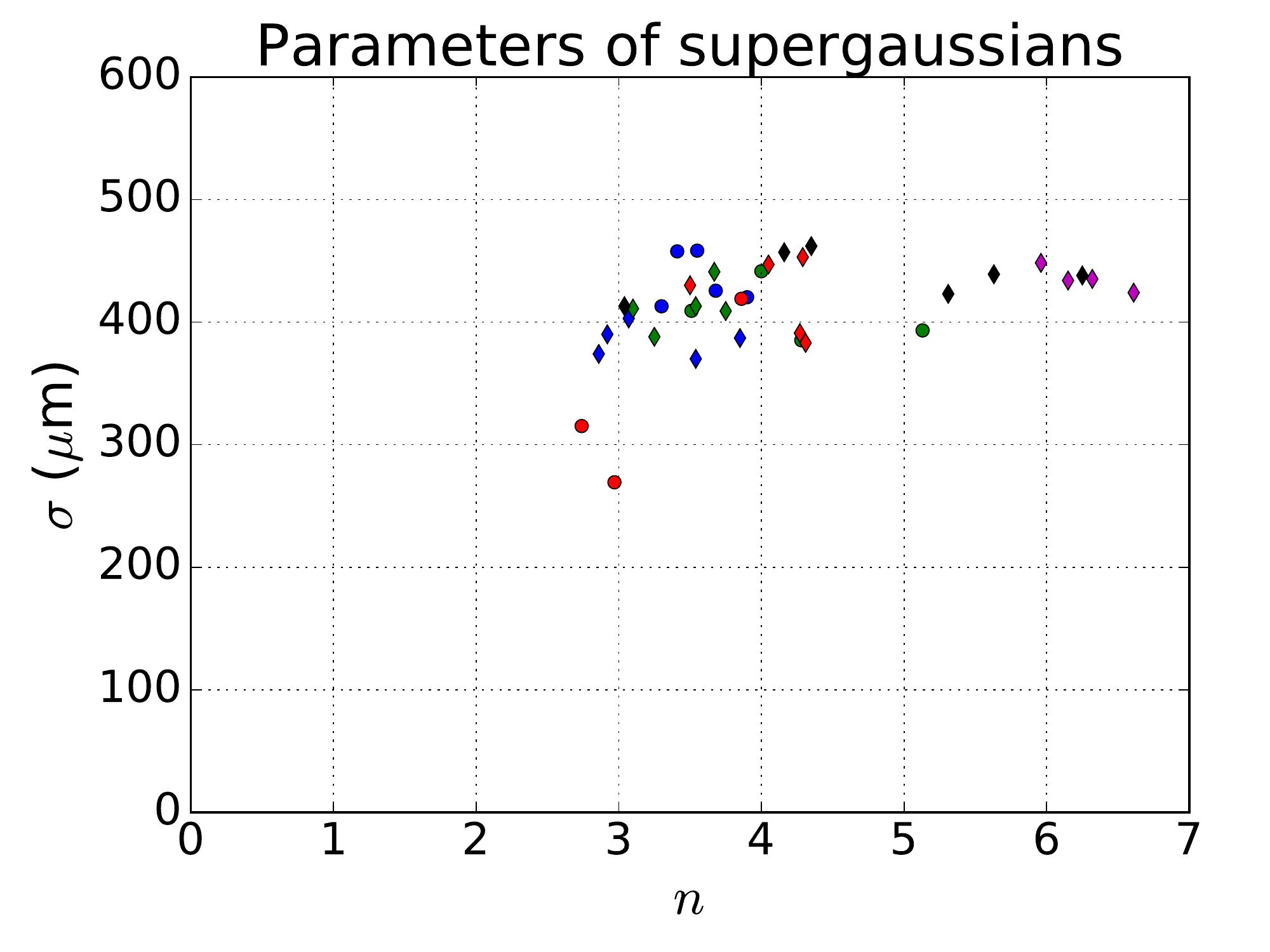} 
\caption{(Color online) Plot of the optimal parameters $n$ and $\sigma$ for the 37 selected XRFC-pictures (cf. Tab.\ref{tab:table1}). The half length $\sigma$ of the cord, intersection of the tube axis with the BL spot of the spot, is concentrated around 400 $\mu$m which is characteristic of the half diameter of the spot meaning that in most cases the center nearly coincides with the tube axis (apart from two points with smaller half length) and $n$ extends from 3 to 6 with a large population around 4. Same shape and colors of points means images from the same shot.}\label{fig:nsigma}
\end{figure}

The procedure would not have given the same result if points from (a.) were cast away. These points, even if there is a few of them as compared to points in (u.s.), are important for they contain the information on the extension of the supergaussian. Without these points, the fit over (u.s.) would have given a correct value of the exponent $n$ but a loose value of $\sigma$. This is the reason why having an undoped piston was a chance as mentioned at the end of section \ref{sec:st}. 

\par 

To summarize, these four steps have allowed to find a good guess of the backlighter intensity profile, namely the parameter $I_0$, $z_c$, $\sigma$ and $n$ in function (\ref{ffit}), out of the raw experimental intensity profiles $I_\mathrm{raw}(z)$. Since, transmitted intensities are proportional to incident intensities, the quantity $I_\mathrm{raw}(z)/f(z, I_0, z_c, \sigma, n)$, ratio of the experimental transmitted intensity over the extrapolated incident intensity, called the experimental renormalized intensity profile hereafter, can readily be compared to a simulation of the transmission which assumes a uniform source of x-rays from the backlighter. 

\subsection{Comparison of experimental renormalized intensity profiles with simulations}

The selected raw intensity profiles for each shot of both campaigns 2013 and 2015 and the results of the previously described procedure is summarized in Fig.\ref{fig:radprof} where plots of renormalized intensity profiles are underneath plots of raw intensity profiles. 

\par

By construction, as can be seen from this figure, renormalized intensity profiles superimpose well over the unshocked area for all radiographic pictures in a given shot. What is more surprising from an image processing point of view, but quite comforting from a physical point of view, is the fact that these renormalized profiles collapse also very well over the shocked region (s.) whereas this region was not involved in the whole fitting procedure. This was not obvious from the start because for each raw intensity profile a different fitted function $f$ is found, with different values of $n$ and $\sigma$ (because the backlighter spot moves slightly during the pulse and is not bound to be centered on the axis of the tube), and the ratio of the two, the so called renormalized intensity profile, could have been affected by this procedure. It does prove that, for a given shot, the transmission over the shocked and unshocked regions does not change over the time of the backlighter pulse. This transmission can then be compared to the simulated transmission with DIANE (Fig.\ref{fig:rad2} and solid black lines on Fig. \ref{fig:radprof}). The collapsed experimental renormalized intensity profiles match quite well with all simulated DIANE normalized intensity profiles, apart from shot 78807 where one can clearly see a shift in the shock position attributed to the equation of state used to simulate the aerogel (this problem is deferred to the next section \ref{sec:sop}).   

\par

In the ablated region (a.), on the other hand, one cannot jump to the same conclusion. Indeed, in most cases, the renormalized intensity profiles superimpose well but in some case they do not and the level of renormalized intensity may vary up to a factor 3 within the 750 ps of the backlighter pulse. 
\par 

\begin{figure}[h] 
\includegraphics[width=8cm]{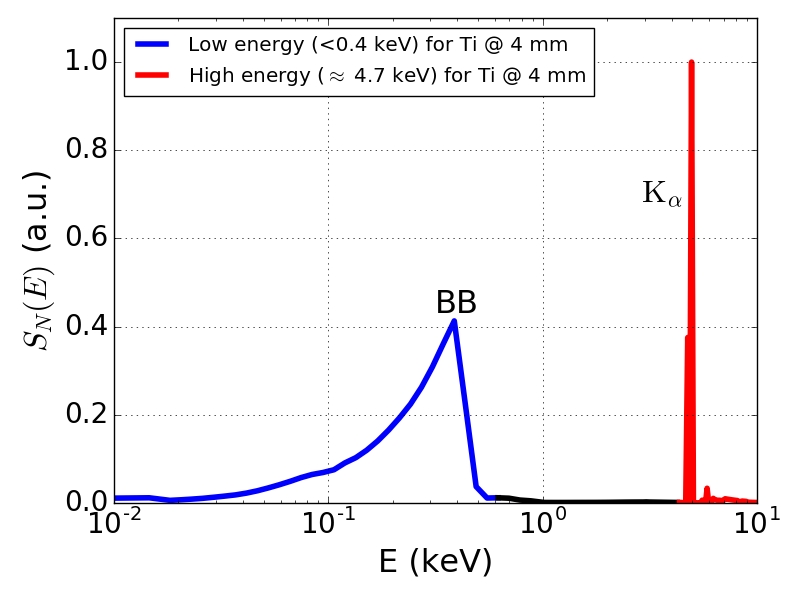} 
\caption{Density spectrum, $S_N(E)$, of a titanium backlighter at t=0.9 ns of the nanosecond pulse. The $K_\alpha$ emission line at 4.7 keV for titanium (on the right) is well  separated from the ``truncated'' black-body spectrum below 0.4 keV. It is this last contribution that makes the renormalized intensity contribution of the ablated region to vary so quickly during the laser pulse of the backlighter.}\label{fig:specti}
\end{figure}

\begin{figure}[h] 
\includegraphics[width=8cm]{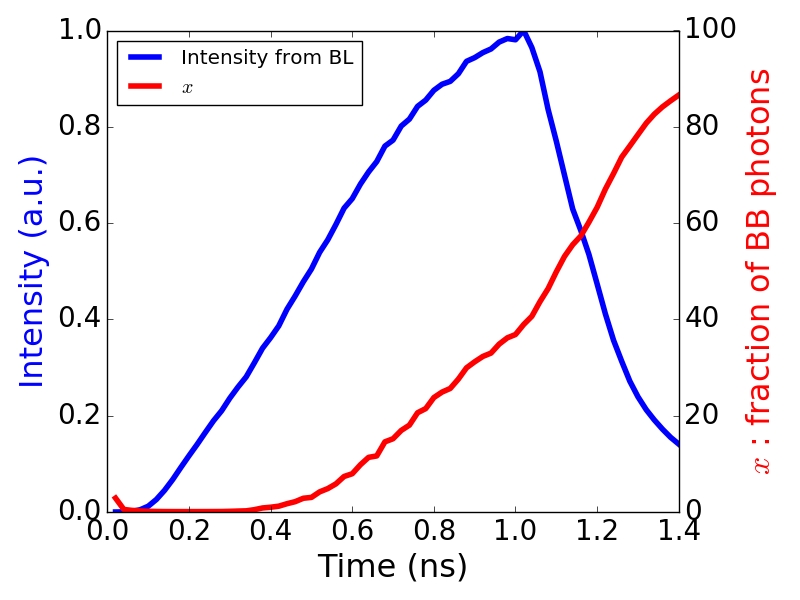} 
\caption{(Color online) The blue curve is the intensity (in arbitrary units) of radiations from the backlighter (BL) versus time as seen from the shock-tube. After 1 ns, it decreases rapidly because laser beams are turned off. The red curve represents, on the same time scale, the fraction of black-body (BB) photons emitted and transmitted by the backlighter over the total number of photons, including the $K_\alpha$ photons. That fraction increases steadily from 0 to 40 $\%$ by the end of the pulse and continues to grow after.}\label{fig:frac}
\end{figure}

This might stem from the fact that the backlighter power spectrum, $S_\mathrm{E}(E,t)$ (where $\int S_\mathrm{E}\,\mathrm{d}E$ is the integrated energy received per unit time), changes during the pulse. At early time, most photons emitted have peaked energy around the emission energy $E_\alpha$ keV ($E_\alpha=4.3$ keV for scandium, 4.7 for titanium, etc.). Therefore $S_\mathrm{E}(E,t)\propto \delta(E-E_\mathrm{\alpha})$ at early time. As time goes by, the power spectrum remains peaked around $E_\alpha$ but a more and more important fraction of low energy photons (below 0.4 keV) may build up due to the backlighter black-body emission (see Fig.\ref{fig:specti}). The black-body emission is ``truncated'' above 0.4 keV (in the case of titanium) because of the backlighter L-band absorption. What is even more important than power spectrum, from an XRFC point of view, is the density spectrum (number of photons per energy bins, $S_N(E,t)=S_\mathrm{E}(E,t)/E$, where $\int S_\mathrm{N}\,\mathrm{d}E$ is the integrated number of photons received per unit time). Indeed, the XRFC detects number of photons. From this perspective, the weight of low energy photons is enforced. Simulations of the backlighter show that the fraction of low energy photons increases steadily throughout the nanosecond pulse of the backlighter (see Fig.\ref{fig:frac}). Even if their energy balance is negligible to a certain extent, their population is not and is even shoulder to shoulder with that of $K_\alpha$ photons by the end of the pulse: low energy photons count for 40$\%$ at that time. 

\par 

The hydrodynamics of the backlighter is of the essence : all this is very sensitive to non-local thermodynamic equilibrium (NLTE) models. It must be emphasized that the use of an NLTE model tends to generate smaller electron temperature compared to an LTE situation. The ion temperature is subsequently smaller and so is the velocity of the shock transmitted in the backlighter by the laser pulse. The consequence is important on the intensity and timing of the rarefaction reflecting off the free side of the backlighter. Indeed, when the shock reaches that free side, a rarefaction propagates back into the backlighter and decompresses it. The density of the backlighter material drops more or less rapidly (depending on the NLTE model) allowing (or not) both spectral components ($K_\alpha$ and black-body) to be transmitted through the BL instead of one ($K_\alpha$).

\par

But what does it have to do with the intensity variation observed in the ablated region ? 

\par 

\begin{figure}[h] 
\includegraphics[width=8cm]{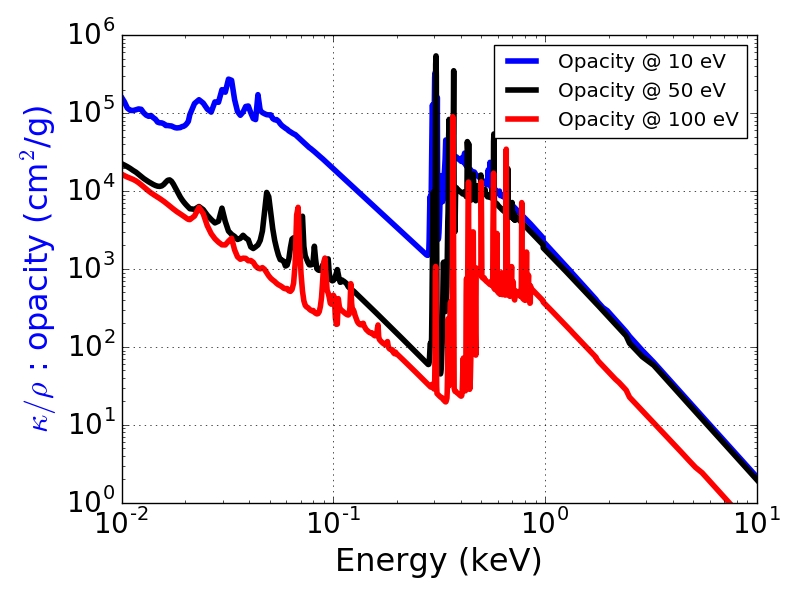} 
\caption{(Color online) Opacity of DVB aerogel at 0.05 g/cm$^3$ at three different temperatures: $T$=10 eV in blue, 50 eV in black and 100 eV in red. The strong discontinuity with a forest of emission lines in all three curves at photon energy of approximately $E$=0.3 keV corresponds to the K-shell of carbon. On one hand, between $T$=10 and 50 eV and for $E<$0.3 keV, the opacity drops by 1.5 orders of magnitude but stands still for $T>$50 eV. On the other hand, between $T$=10 and 50 eV and for $E>$0.3 keV, the opacity does not change, whereas it drops by less than one order of magnitude when $T$ increases from 50 to 100 eV.}\label{fig:opatemp}
\end{figure}

Let us take the example of shot 78804 (Ti backlighter, with density and temperature ploted at the bottom of Fig.\ref{fig:TIRO}). In the unshocked (u.s.) region, density is $\rho_0$=0.3 g/cm$^3$ with temperatures not exceeding 10 eV (see Fig.\ref{fig:TIRO}). In the ablated region, on the other hand, densities are closer to $\rho_a\approx 0.05$ g/cm$^3$ with temperatures above 60 eV. Now (cf. Fig.\ref{fig:opatemp}), the opacity of the aerogel varies significantly between \mbox{$T$=10 eV} and 100 eV. For photons with energy $E>0.3$ keV, the opacity remains constant up to \mbox{$T$=50 eV} and drops by a factor 6 between \mbox{$T$=50 eV} and 100 eV. It is the other way around for photons with energy $E<0.3$ keV. The opacity drops by a factor 30 between \mbox{$T$=10 eV} and 50 eV and remains approximately constant between \mbox{$T$=50 eV} and 100 eV. 

\par 

For the sake of clarity, let's assume that the backlighter spectrum is approximately made of two monochromatic components: one at $E=E_\alpha$ (the energy of the $K_\alpha$ emission line) and one at $E=E_{\mathrm{BB}}$ (where $E_\mathrm{BB}$ is the average of $E$ weighed by the black-body (BB) part of the $S_N$ distribution of Fig.\ref{fig:specti}). In the case of titanium, $E_\alpha$=4.7 keV and $E_\mathrm{BB}\approx$ 0.3 keV. If $x$ is the fraction of black-body radiation, then the BB photons flux going through the tube is $x\,\Phi_0$, where $\Phi_0$ is the total flux of photons from the backlighter ($E_\mathrm{BB}$ plus $E_\alpha$ contributions), and $(1-x)\,\Phi_0$ is the $K_\alpha$ photons flux. Therefore, the flux transmitted through the unshocked region, $\Phi^{t}_\mathrm{us}$, and the flux transmitted through the ablated region, $\Phi^{t}_\mathrm{a}$, are
\begin{align}
 \frac{\Phi^{t}_\mathrm{us}}{\Phi_0}&=(1-x)\,e^{-\frac{\kappa}{\rho}(E_\alpha,T_0)\,\rho_0\,\ell}+x\,e^{-\frac{\kappa}{\rho}(E_\mathrm{BB},T_0)\,\rho_0\,\ell},\\
 \frac{\Phi^{t}_\mathrm{a}}{\Phi_0}&=(1-x)\,e^{-\frac{\kappa}{\rho}(E_\alpha,T_a)\,\rho_a\,\ell}+x\,e^{-\frac{\kappa}{\rho}(E_\mathrm{BB},T_a)\,\rho_a\,\ell}.
\end{align} the ratio $\Phi^{t}_\mathrm{a}/\Phi^{t}_\mathrm{us}$ gives us the value of the renormalized intensity to be expected in the ablated region. When $x$= 0, the ratio is $\approx e^{+\frac{\kappa}{\rho}(E_\alpha,T_0)\,\rho_0\,\ell}$ which is $\approx$ 1.7 for titanium and when $x$=0.4 (anticipated by BL simulations) the ratio is close to 3 which is consistent with the renormalized intensity profiles of this shot Fig.\ref{fig:radprof}. For the other shots, values are different because backlighters are different. 

\par

In a nutshell, the shocked and unshocked region only let through $K_\alpha$ photons whereas the ablated region, thiner, can let through both $K_\alpha$ and $\mathrm{BB}$ photons. Since the shocked and unshocked regions let only one component of the radiation go through, their contrast, on the renormalized intensity profile, does not vary during the course of the pulse. It does not depend on the intensity of the incident $K_\alpha$ line (it cancels out on the renormalized intensity profile). The contrast only depends upon the thermodynamic conditions of both states (shocked and unshocked) that do not vary much during the backlighter pulse. On the contrary, the ablated region let both radiations ($K_\alpha$ and $BB$) go through and since the BB component varies rapidly during the pulse, so does the contrast of the renormalized intensity of the ablated region.

\par 

%
%
\subsection{\label{sec:sop} Comparison of experimental SOP images with trajectory inferred from simulation}

\begin{figure}[h] 
\fbox{\includegraphics[width=3.9cm]{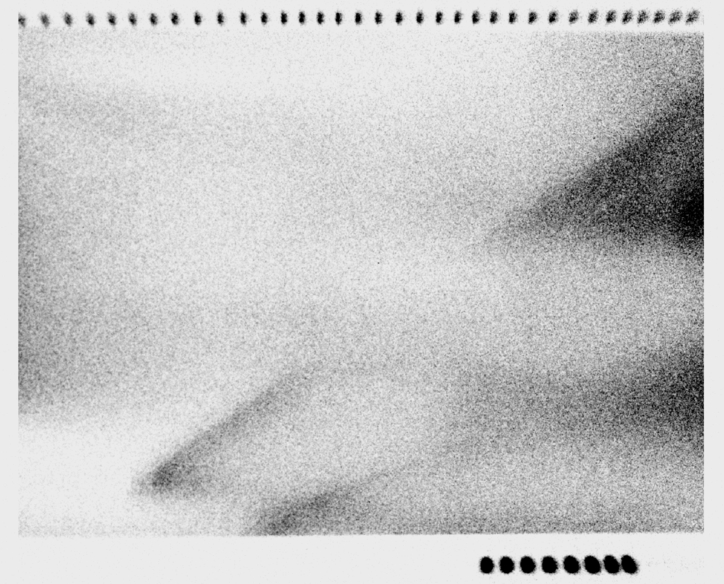}}\\
\vspace{.1cm}
\fbox{\includegraphics[width=3.9cm]{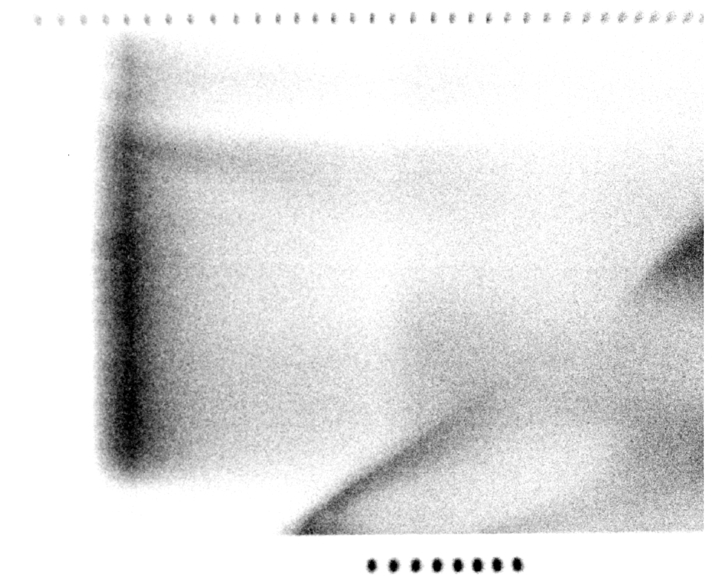}}\hspace{0.2cm}\fbox{\includegraphics[width=3.9cm]{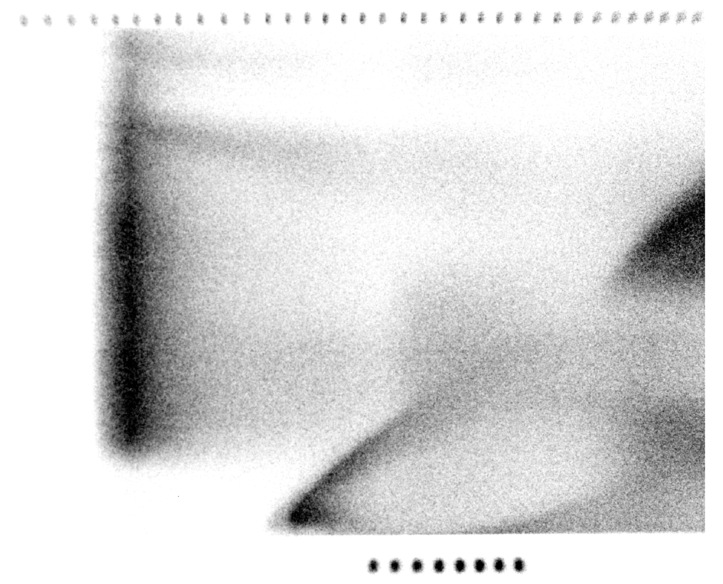}}
\caption{SOP images of shot 70990 (top), 70992 and 70993 (bottom). On the bottom figure, the laser pulse is clearly visible as a dark vertical strip, two clicks wide on the horizontal (time) axis. The two nearly horizontal dark shapes at the top correspond to features at the end of the tube. The time sweep on the top SOP image started later which explains why the vertical dark strip is not visible. One of the dark shape is visible.}\label{fig:sop2013}
\end{figure}

\begin{figure}[h] 
\fbox{\includegraphics[width=3.9cm]{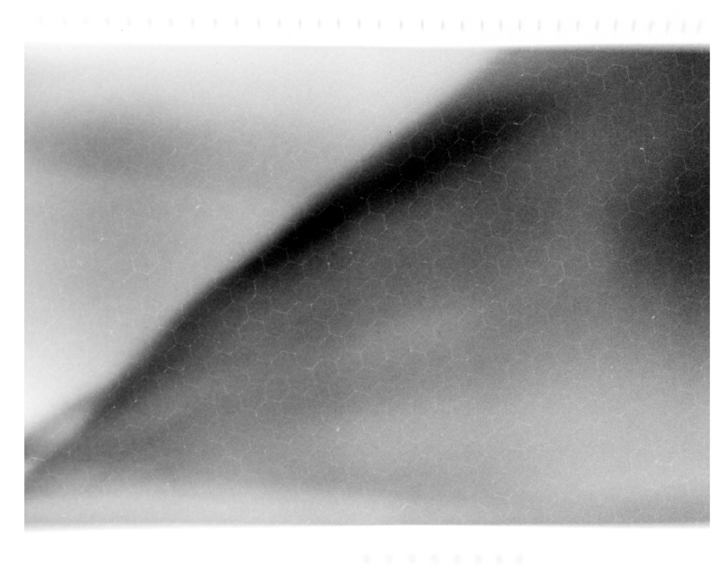}}\hspace{0.2cm}\fbox{\includegraphics[width=3.9cm]{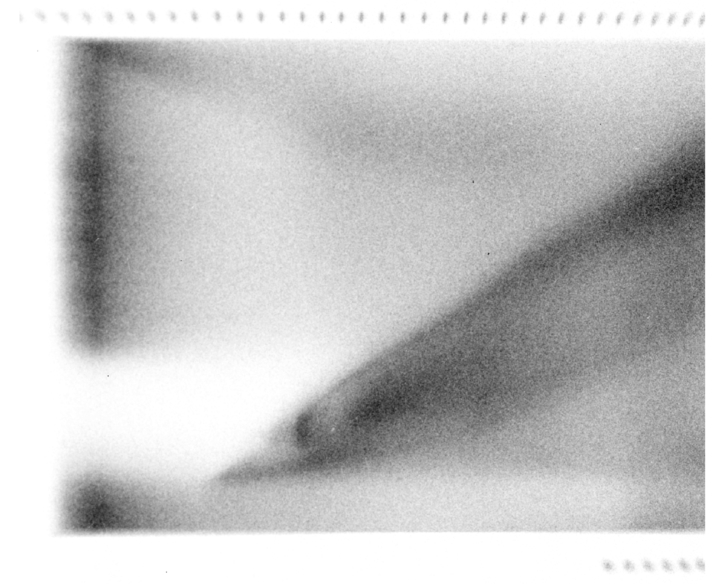}}\\
\vspace{.1cm}
\fbox{\includegraphics[width=3.9cm]{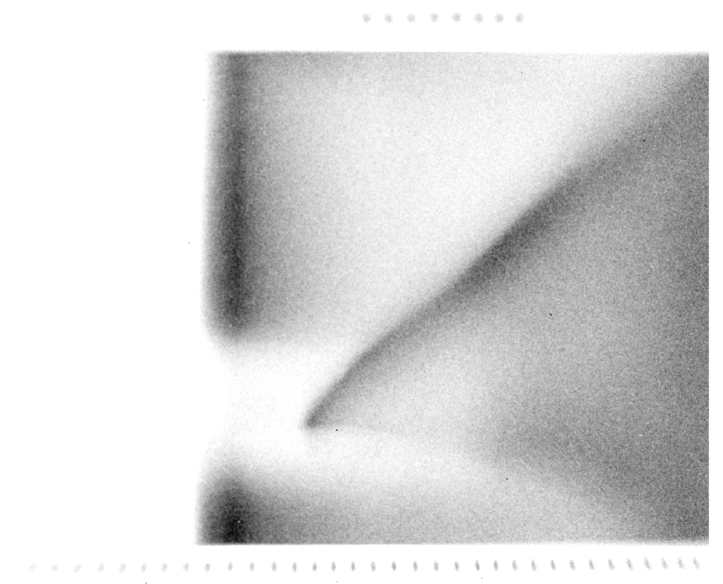}}\hspace{0.2cm}\fbox{\includegraphics[width=3.9cm]{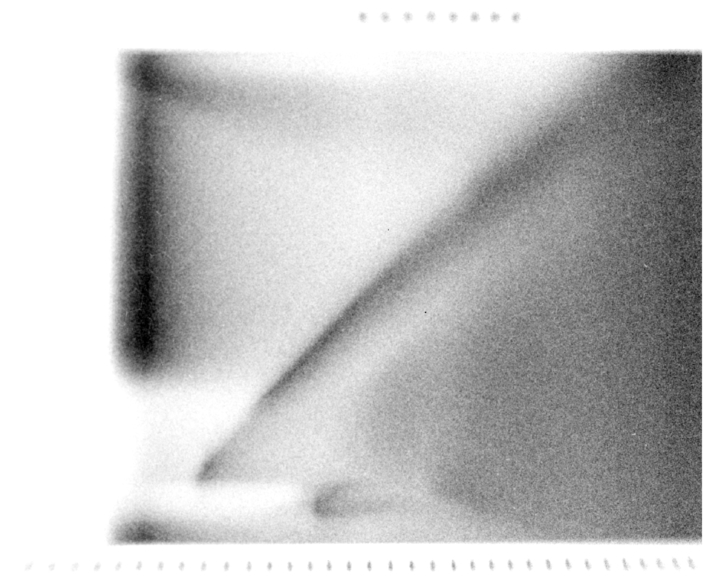}}
\caption{The top figures correspond to SOP images of shots 78804 and 78806 (simple-flux with 300 $\mu$m piston in PI), the bottom figures correspond to shots 78805 and 78807 (simple-flux without piston).}\label{fig:sop2015_1}
\end{figure}

\begin{figure}[h] 
\fbox{\includegraphics[width=3.9cm]{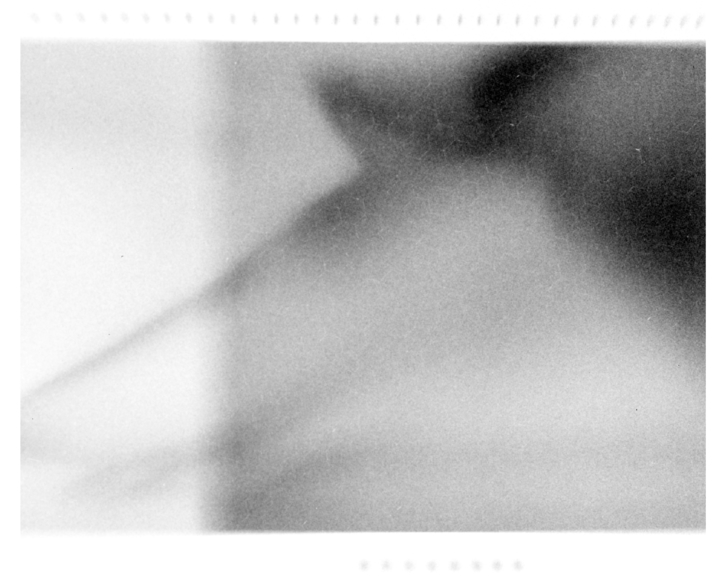}}\hspace{0.2cm}\fbox{\includegraphics[width=3.9cm]{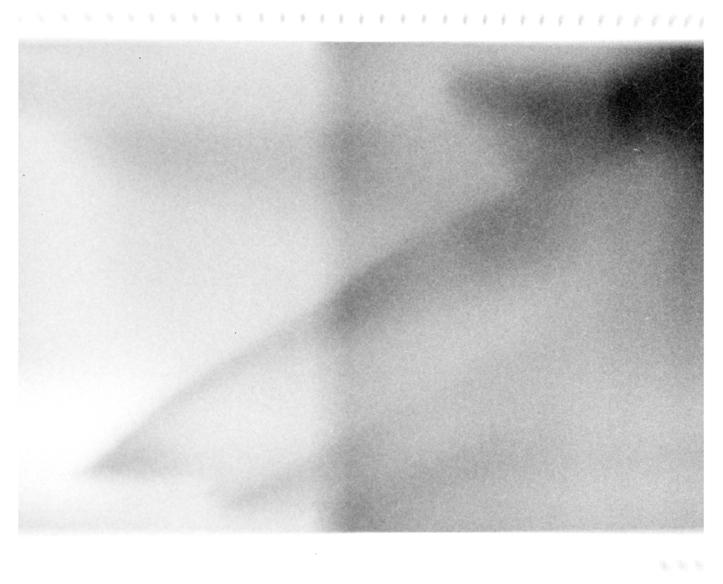}}\\
\vspace{.1cm}
\fbox{\includegraphics[width=3.9cm]{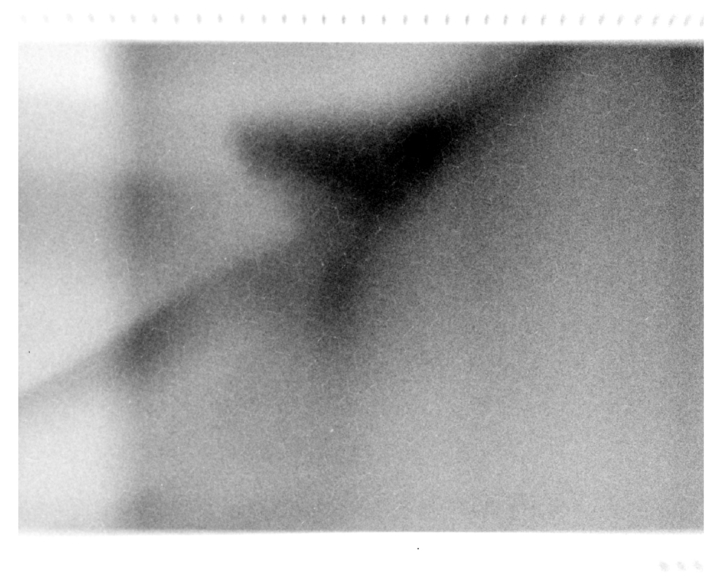}}\hspace{0.2cm}\fbox{\includegraphics[width=3.9cm]{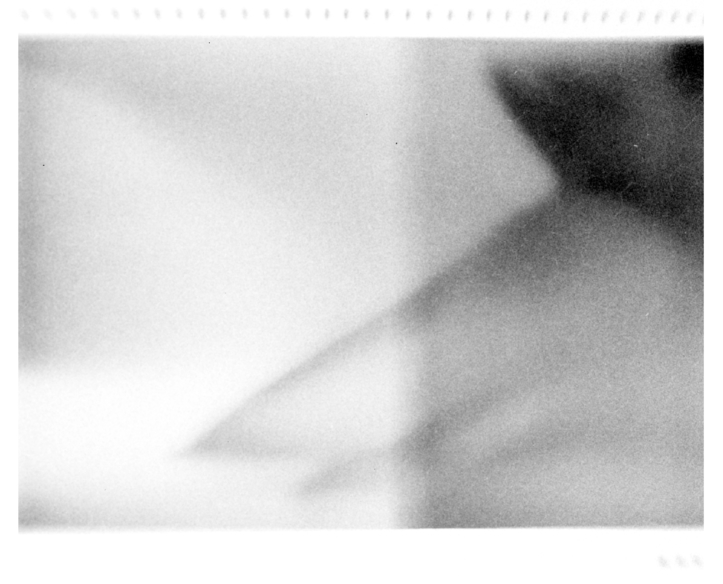}}
\caption{These SOP images correspond (from left to right and top to bottom) to shots 78808, 78809, 78810 and 78811 (double-flux). The time sweep started well after the laser pulse which explains why it is not visble on any of these images. There is, however, on each image, a vertical fuzzy transition from a light area (on the left) to a dark area (on the right) which correspond to the laser pulse of the second halfraum (at 9.5 ns).}\label{fig:sop2015_2}
\end{figure}

The self emission of the tube is collected inside the target chamber by a f/3.3 telescope and is relayed outside through a series of optics to the so called SOP cabinet. The light goes through a long-pass filter that only let through radiations with wavelength greater than 590 nm. The light is brought to the entrance of the Rochester Optical Streack System (ROSS) through another series of optics and filters. A dove prism enables to rotate the image of the tube on the entrance slit to be able to resolve space on a slit (1D image) and time to form a 2D image. Inside the optical streaked system, the self emission is focused on a photocathode turning light into electrons that are accelerated into a phosphor screen. The light emitted by that screen is transported by optical fiber to individual pixels on a 1100$\times$1100 CCD camera. On the ROSS, the sweeping time was set to 17 ns.  A neutral density (ND) filter was placed in front of the slit (different ND where used on both campaign, see table.\ref{tab:table1}) to limit the saturation of the diagnostic (limitation of the photocurrent). 

\par

Each SOP image, on figs.\ref{fig:sop2013}, \ref{fig:sop2015_1} and \ref{fig:sop2015_2}, spans over 17 ns along the horizontal axis and over 800 $\mu$m of the tube along the vertical axis. On each SOP image, at the top, or at the bottom, a series of 35 dots are fiducial time marker equally spaced by 548 ps. On the opposite side, 8 dots are fiducial marker equally spaced by 548 ps and starting when the laser pulse of the backlighter was turned on. On many of these images, on can see the 1 ns pulse as a long dark strip throughout the tube (cf. Fig.(\ref{fig:sop2013}) for instance). The time extension of these strips on the SOP images corresponds exactly to 1 ns. On other images, one can see the end of the tube which allows to have an absolute positioning along the tube. The scale is set by the fact that, on the focal plan of the f/3.3 telescope, with a magnification of $\times$4, the slit length corresponds to 800 $\mu$m (the correction due to the parallax of the TIM5, supporting the telescope for the SOP, which is off the normal of the P6-P7 axis by 10.8 degree is negligible). The slit width was fully open to let a maximum of light go through.

\begin{figure*}[t] 
\includegraphics[width=9cm]{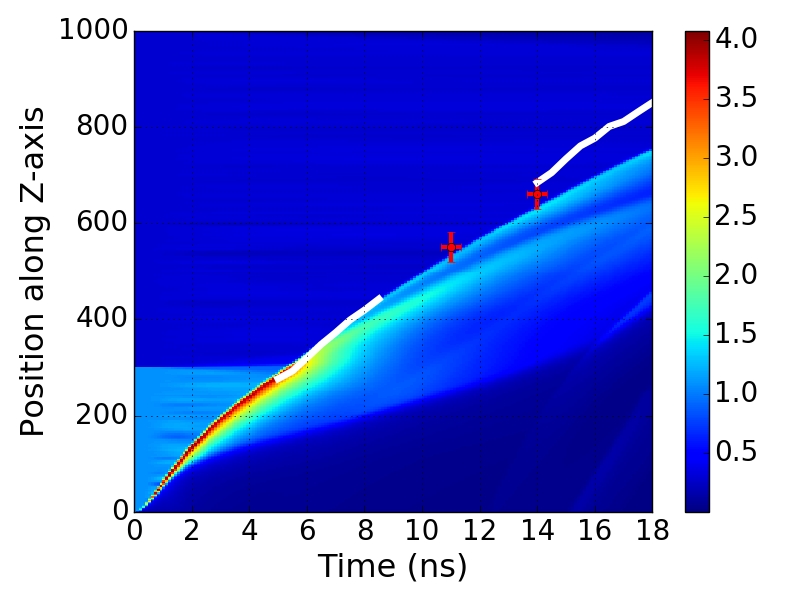}\includegraphics[width=9cm]{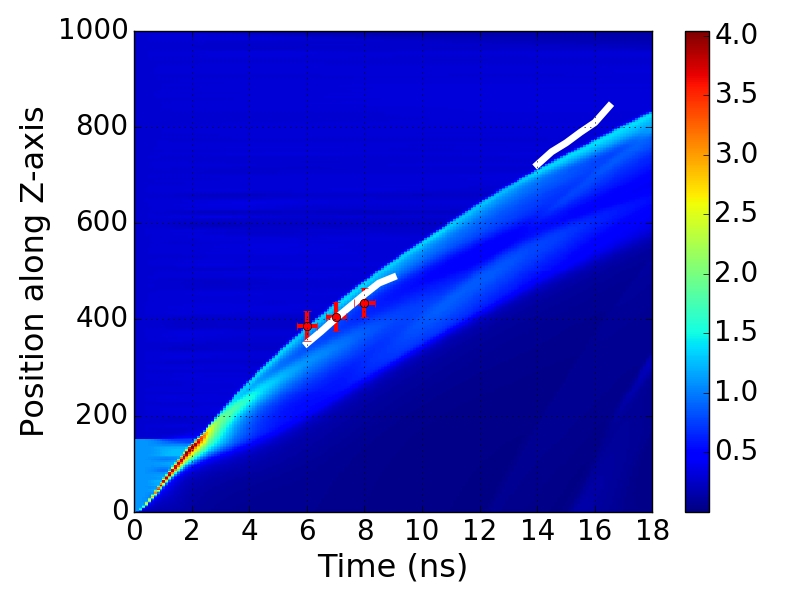}
\includegraphics[width=9cm]{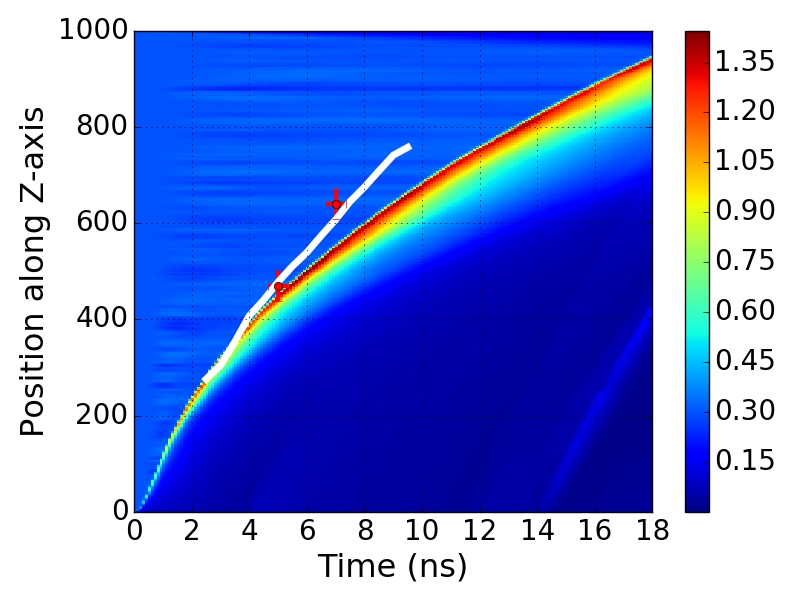}\includegraphics[width=9cm]{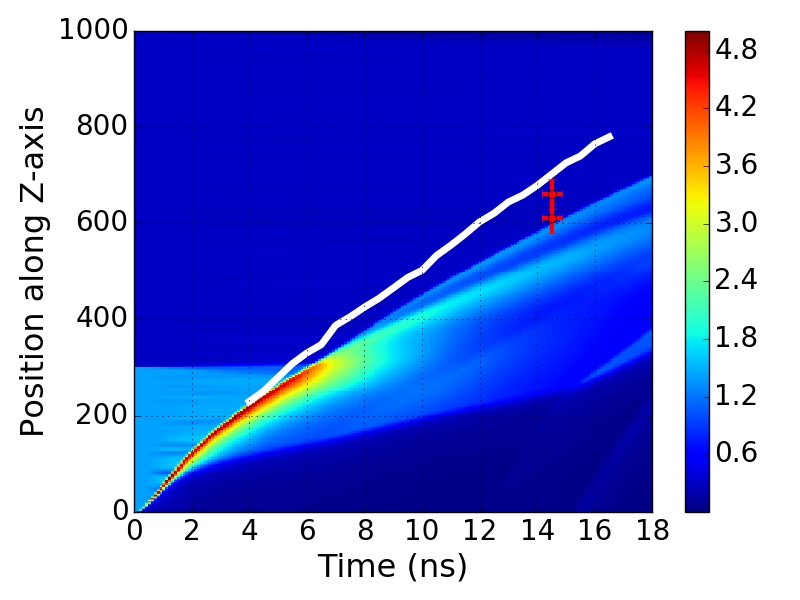}
\caption{(Color online) Color plot of computed density (space, between $z$=0 and 1000 $\mu$m, on the vertical and time on the horizontal axis). The red points with error bars correspond to the experimental position of the shock from XRFC, $z_s$ on tab. \ref{tab:table1}. The white curves are drawn from the SOP images and correspond to the experimental location of the shock versus time. From left to right and top to bottom are : (a) 300 $\mu$m piston in polystyren (70989, 70990), (b) 150 $\mu$m piston in polystyren (70991 to 70995), (c) no piston, only DVB aerogel (78805, 78807) and (d) 300 $\mu$m piston in polyimid (78804, 78806)}\label{fig:pltsop}
\end{figure*}

%

\par

The shock positions at all time have been read on the SOP images and plotted in white on Fig. \ref{fig:pltsop} against the few positions noted down from radiographic images of the same series and against position of the shock from the relevant simulation. The agreement is overall correct but the difference widens at late time. 

\par 

The SOP diagnostic is always difficult to interpret because it observes visible light that can be due to the black-body radiation (since in the Rayleigh-Jeans part of the spectrum, for $h\,\nu\ll k\,T$, the intensity $I(\nu, T)\propto\nu^2\,T$), its primary goal, that works well when observing a shock head-on \cite{}, but side-on, we can invoke other contributions such as fluorescence of the plastic due to the photo-ionization and rearrangement of the electron shells, that can be ahead of the shock, Rayleigh diffusion that can also be ahead of the shock. The trajectories of shocks inferred from SOP images were assumed to be the trajectories of the maximum of the intensity on the SOP, that is to say, the trajectories of the upper dark trace from bottom left to top right on any of the SOP images. That assumption is certainly not accurate to get the correct absolute position of the shock but it is certainly a correct relative marker : the offset between the location of the maximum intensity on the SOP images and the actual location of the shock must remain constant for a shock whose thermodynamical state does not vary too much. Thus, a side-on SOP allows to assess the velocity of the shock and, another way to put it is to say that the trajectory from the SOP must be parallel to the actual trajectory. As can be seen on Fig. \ref{fig:pltsop}, trajectories of shocks deduced from SOP (in white) intercept the red dots corresponding to the XRFC positions meaning experimental data from SOP and XRFC are coherent. Clearly, SOP trajectories deviate from simulated trajectories at late time. Experimental shock velocities seem higher than simulated trajectories especially for shot 78805 and 78807 (with no piston, $d_\mathrm{ab}=0$). 

\par

\begin{figure}[h] 
\includegraphics[width=8cm]{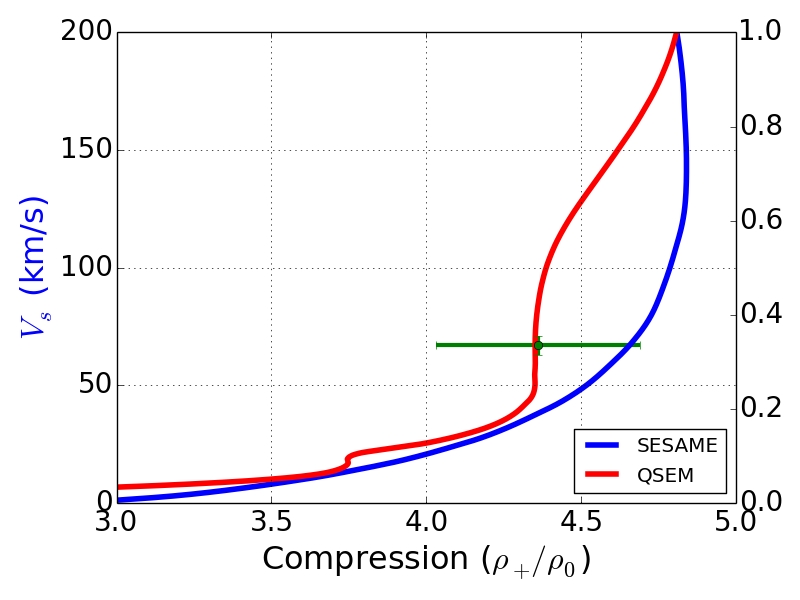}
\includegraphics[width=8cm]{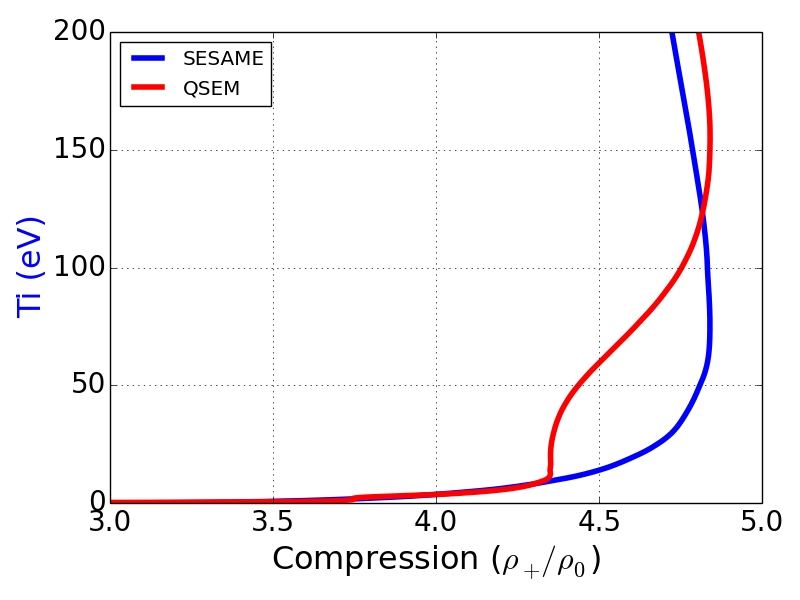}
\caption{Principal Hugoniot of CH with initial state $\rho_0=$ 0.3 g/cm$^3$ and $Ti=$ 0.025 eV for SESAME and QSEM in Shock-velocity versus compression representation top) and ion temperature versus compression (bottom). The green point is extracted from the experimental data (the difficulty relies in determining the density of the shocked state). The QSEM equation of state captures that odd point from the SESAME equation of state perspective.}\label{fig:vitro}
\end{figure}

In theory, from the renormalized intensity profile (out of XRFC images) one can deduce the density ratio on both side of the shock (after an Abel transformation). It is enough information to deduce the pressure jump right after the shock front in the shocked material. Let us call $\rho_-$ and $\rho_+$, respectively, the density of the unshocked and shocked aerogel on both sides of the shock front, $P_-$ and $P_+$, the respective pressures and $V_s$ the velocity of the shock in the laboratory frame. Using the conservation of mass ($\rho_-\, V_s=\rho_+\,v$) and momentum ($\rho_-\, V_s^2+P_-=\rho_+\, v^2+P_+$) on both side ($\pm$) of the shock, one can eliminate $v$, which is the shocked velocity in the shock reference frame. The shocked pressure is readily found as
\begin{align}
 P_+\approx \frac{\rho_-}{\rho_+}\,(\rho_+-\rho_-)\,V_s^2.\label{eqpplus}
\end{align} 

\par 

In our particular experiments, since shocks are curved, the intensity front profiles between unshocked and shocked region is not as abrupt as it would appear on a density profile for all shots (apart from shots 78805 and 78807). This is due to the fact that intensity profiles integrate density along the optical path of photons that cross the curved shock surface from the backlighter to the XRFC. Therefore, it also integrates the complex density profile behind the shock (in the shocked region). Owing to the presence of the piston in the shocked region of all shots (but 78805 and 78807), the density is wobbling rapidly behind the shock and intensity profiles around the shock cannot be easily used to extract density ratios on both side of the shock. Luckily, for shots 78805 and 78807, without piston, in pure aerogel, the profile behind the shock looks like an ideal blast wave (see Fig. \ref{fig:TIRO}) and the simulated profiles superimpose well to the experimental profiles, meaning experimental and simulated density profiles agree well for these shots. From Fig. \ref{fig:TIRO}, one deduces that the simulated shocked density of $\rho_+=1.31 \pm 0.1$, corresponding to a compression of $4.36 \pm 0.33$, agrees with the experimental data but the simulated (50 km/s) and experimental (67 km/s) shock velocities disagree significantly. 

\par

The principal Hugoniot (in the representation shock velocity versus compression) of the SESAME equation of state used for our simulation is described on Fig. \ref{fig:vitro} along with the QSEM \cite{huser2015, huser2016} equation of state created specially for our aerogel. This more sophisticated equation of state, using the so called quantum semiempirical model (QSEM) \cite{huser2015, huser2016}, was compared graphically but not yet tested in these simulations. It is an average atom-model where the electron part comes from a fully quantum calculation and the cold curve is provided by experimental data or calculated by quantum molecular dynamics (QMD), the ion part being the classical Cowan ion equation of state \cite{cowan}. The experimental shocked point in green in Fig. \ref{fig:vitro}, with $\rho_+$=1.31$\pm$ 0.1 g/cm$^3$, $V_s$= 67 $\pm$ 4 km/s, corresponding to a shocked pressure $P_+$= 10 $\pm$ 1 Mbar using eq.(\ref{eqpplus}), gives a clear advantage to the QSEM equation of state. It is the reason why the shock is faster experientially in shot 78807 (see Fig. \ref{fig:radprof}). Clearly, the SESAME equation of state gives smaller shock velocity for compressions above 4.3 at an equivalent temperature than the QSEM. It suggests that the shocked state in the aerogel is in the isochore part of the QSEM.

%

%
%

\section{Conclusion}

The radiographic image processing tested on the {\it simple-flux} and {\it double-flux} shock-tube experiment at OMEGA is very promising. It has proved rather efficient at removing the backlighter background providing the radiographic experimental profiles a way to be compared to simulated profiles. The experimental {\it renormalized intensity profiles} collapse well on each other. These intensity profiles can be used to get information on the density profile of the material inside the shock-tube, in the shocked region but also in the ablated region. It has provided us with an important Hugoniot point of our CH aerogel which pleads the cause of an isochore branch of the principal Hugoniot at compression $\approx 4.3$ as predicted by a QSEM equation of state. It also brought to the fore the possibility of a low energy contribution from the backlighter spectrum to the rapid variation of the ablated region relative transmission in radiographic images. In our simulations, this effect is very sensitive to the NLTE model used. This effect could be engineered in a future dedicated experiments to discriminate between these models.

%
%
%
%

\begin{acknowledgments}
The authors would like to acknowledge the use of FIJI \cite{fiji} (FIJI Is Just ImageJ) to post-process the images. They also express their sincere gratitude to Gwenael Salin who computed the QSEM equation of state for the aerogel. 
\end{acknowledgments}

\bibliographystyle{plain}

\end{document}